\documentclass[10pt]{article}
\usepackage{color}
\usepackage{amssymb} 
\usepackage{bm} 
\usepackage{amsthm}
\usepackage{amsmath} 
\usepackage[percent]{overpic} 
\usepackage{graphicx}
\newcounter{defn}
\newcounter{axm}
\newcounter{lm}
\newcounter{lm2}
\newcounter{thm}
\newcounter{thm2}
\newcounter{pst}
\newcounter{cor}
\newtheorem{definition1}[defn]{Definition}
\newtheorem{definition2}[defn]{Definition}
\newtheorem{definition3}[defn]{Definition}
\newtheorem{definition2.5}[defn]{Definition}
\newtheorem{definition4}[defn]{Definition}
\newtheorem{definition5}[defn]{Definition}
\newtheorem{axinf}[axm]{Axiom}
\newtheorem{axmic}[axm]{Axiom}
\newtheorem{axsecond}[axm]{Axiom}
\newtheorem{lmfunc}[lm]{Lemma}
\newtheorem{lmremove}[lm]{Lemma}
\newtheorem{lmfuncApp}[lm2]{Lemma}
\newtheorem{lmremoveApp}[lm2]{Lemma}
\newtheorem{lmremoveREDApp}[lm2]{Lemma}
\newtheorem{lmMoreCondApp}[lm2]{Lemma}
\newtheorem{thmconst}[thm]{Theorem}
\newtheorem{thmconstApp}[thm2]{Theorem}
\newtheorem{thmentropy}[thm]{Theorem}
\newtheorem{thmentropyApp}[thm2]{Theorem}
\newtheorem{thmMicroMacro}[thm]{Theorem}
\newtheorem{thmMicroMacroApp}[thm2]{Theorem}
\newtheorem{thmLandauer}[thm]{Theorem}
\newtheorem{thmLandauerApp}[thm2]{Theorem}
\newtheorem{thmSecondLaw}[thm]{Theorem}
\newtheorem{thmSecondLawApp}[thm2]{Theorem}
\newtheorem{thmindiscernibility}[thm]{Theorem}
\newtheorem{thmindiscernibilityApp}[thm2]{Theorem}
\newtheorem{thmMutual}[thm]{Theorem}
\newtheorem{thmMutualApp}[thm2]{Theorem}
\newtheorem{thmAdditivity}[thm]{Theorem}
\newtheorem{thmAdditivityApp}[thm2]{Theorem}
\newtheorem{thmClairvoyance}[thm]{Theorem}
\newtheorem{thmClairvoyanceApp}[thm2]{Theorem}
\newtheorem{pstmain}[pst]{Postulate}
\newtheorem{corLandauer}[cor]{Corollary}
\newtheorem{corSimpler}[cor]{Corollary}
\newtheorem{corTheorem}[cor]{Corollary}

\begin{document}

\title{Daemon computers versus clairvoyant computers: A pure theoretical viewpoint towards energy consumption of computing}
\author{Alireza Ejlali \\ Sharif University of Technology\\ Department of Computer Engineering\\ ejlali@sharif.edu}
\maketitle

\begin{abstract}
Energy consumption of computing has found increasing prominence but the area still suffers from the lack of a consolidated formal theory. In this paper, a theory for the energy consumption of computing is structured as an axiomatic system. The work is pure theoretical, involving theorem proving and mathematical reasoning. It is also interdisciplinary, so that while it targets computing, it involves theoretical physics (thermodynamics and statistical mechanics) and information theory. The theory does not contradict existing theories in theoretical physics and conforms to them as indeed it adopts its axioms from them. Nevertheless, the theory leads to interesting and important conclusions that have not been discussed in previous work. Some of them are: (\emph{i}) Landauer\textquoteright s principle is shown to be a provable theorem provided that a precondition, named macroscopic determinism, holds. (\emph{ii}) It is proved that real randomness (not pseudo randomness) can be used in computing in conjunction with or as an alternative to reversibility to achieve more energy saving. (\emph{iii}) The theory propounds the concept that computers that use real randomness may apparently challenge the second law of thermodynamics. These are computational counterpart to Maxwell\textquoteright s daemon in thermodynamics and hence are named daemon computers. (\emph{iv}) It is proved that if we do not accept the existence of daemon computers (to conform to the second law of thermodynamics), another type of computers, named clairvoyant computers, must exist that can gain information about other physical systems through real randomness. This theorem probably provides a theoretical explanation for strange observations about real randomness made in the global consciousness project at Princeton University.  
\end{abstract}

\section{Introduction} \label{Section:Introduction}
While theoretical computer science is rich in theories of computability and complexity \cite{Sipser}, it has less involvement with energy consumption of computing. However, nowadays, energy consumption has become a prominent factor in computing systems and there are solid evidences \cite{Feynman,Bennett} that unlike what might first come to mind, the minimum energy consumption required for a computation is not proportional to its computational complexity. This implies that current theories in theoretical computer science cannot by themselves address the issue of energy consumption. Indeed, as we will see, it seems that any theory seeking to explain energy consumption of computing has a highly interdisciplinary nature, involving computer science, thermodynamics, statistical mechanics, and information theory.

The basic question is \textquotedblleft Regardless of the way a given computation is implemented (e.g., by means of electronics, photonics, bionics, etc.), what is the minimum value of its energy consumption enforced by the laws of nature?\textquotedblright At present, Landauer\textquoteright s principle \cite{Landauer,Bennett2} is the most notable attempt to answer this question. It states that information erasure is the underlying reason for requiring energy consumption in computation and gives the minimum energy consumed due to erasing a bit. While Landauer provided arguments mainly based on the second law of thermodynamics to justify the principle, he did not provide any formal mathematical proof for it (see \cite{Landauer}). Indeed, following the lack of a formal mathematical proof/theory, the principle has been controversial. There have been some notable critiques of the principle \cite{Earman,Shenker,Norton,Norton2} and again the most notable defenses of the principle \cite{Bennett2,Bennett3} do not establish a formal mathematical theory. There is also a notable experimental study \cite{Hong} that reports observations in favor of the principle. In short, the area still sufferers from the lack of a formal mathematical theory \textemdash ~a theory with specified axioms, well-defined theorems with proofs, and formal mathematical relations. This paper is an effort towards forming such a theory. 

To develop the theory, the axioms have been derived from well-established propositions in thermodynamics and statistical mechanics. On the other hand, information theory has been used as a mathematical tool for describing the theory and developing its equations and relations. Throughout the paper, there are lemmata (listed in Appendix 1) and theorems (listed in Appendix 2), and the distinction between them has been made in such a way that the contribution of the paper mainly lies in the theorems and not in the lemmata, which are pure information-theoretical and have been provided to ease structuring the theory. There is also one postulate in the theory (Section~\ref{Section:Information}) which, unlike the axioms that are derived from the existing context, is proposed in this paper. Indeed, it is called postulate and not axiom to place emphasis on this difference. The postulate has been supported by evidences. Markedly, it has been shown that the postulate is indeed a provable theorem within the axiomatic framework of statistical mechanics and thermodynamics. Also, a section (Section~\ref{Section:Conforming}), including several important theorems, has been dedicated almost entirely to support the postulate.   

The theory has resulted in some interesting and key conclusions:

\begin{itemize}
\item It is proved that Landauer\textquoteright s principle only holds under the condition of macroscopic determinism, which is formally defined in Section~\ref{Section:Energy}. Unlike previous controversial arguments for Landauer\textquoteright s principle, this proof is not based on the second law of thermodynamics. Rather, it is mainly based on the law of conservation of information, which is one of the fundamental laws of nature \cite[Lecture 1]{Susskind}. Furthermore, the proof captures Landauer\textquoteright s principle in the form of a simple information theoretic equation.

\item Landauer\textquoteright s principle has been the main reason behind the use of reversible computing \cite{Frank} to achieve low energy consumption. However, the theory in this paper indicates that randomness might be intentionally exploited to break macroscopic determinism thereby going beyond the energy saving achievable from using reversible computing on its own. It is noteworthy that here randomness, which is formally defined in Section~\ref{Section:Energy}, must be real and pseudo-randomness is of no use.

\item As an interesting case of exploiting randomness for low energy consumption, it has been discussed that NP-complete problems (that comprise an important class of problems with many practical applications \cite{Sipser, Cook}) have a special feature that enables exploiting randomness to reduce the energy consumed in solving the problems. 

\item Following the above item, the theory has the consequence that some computers may exploit randomness in their computation to challenge the second law of thermodynamics. The issue is similar to what is known as Maxwell\textquoteright s daemon \cite{Bennett3, Leff} in statistical mechanics and thermodynamics and hence in this paper these computers are named daemon computers. The theory indicates that randomness plays a central role not only in daemon computers but also in Maxwell\textquoteright s daemon.

\item The second law of thermodynamics is one of the fundamental laws of nature and is generally recognized as an unbreakable law by scientists. To conform to this law, we need to assume that daemon computers cannot exist. But if we assume so, then the theory implies that another type of computers, called clairvoyant computers in this paper, must exist that can gain information about other physical systems through exploiting randomness. This conclusion presumably provides a theoretical explanation for a notable practical observation made in a project at Princeton University called the global consciousness project \cite{GCPP}.
  
\end{itemize}

The issue of energy consumption of computing has some topics in common with some other areas, e.g., physics of information \cite{Zurek}, in terms of involving information theory, thermodynamics and statistical mechanics. Nevertheless, to the best of the author\textquoteright s knowledge, a theory like the one presented in this paper has not been provided in any previous work.

The rest of the paper is organized as follows. Sections 2 and 3 provide the basis of the theory, including several important definitions, axioms, and theorems. Section 4 describes how randomness can be exploited in conjunction with or as an alternative to reversibility to achieve low energy consumption. Specifically, Section 5 addresses the use of randomness to reduce the energy consumption required for solving NP-complete problems. Sections 6 and 7 present and describe the concepts of daemon computers and clairvoyant computers respectively. Section 8 describes how the theory presented in this paper conforms to existing theories in theoretical physics, and finally Section 9 summarizes the paper.

\section{Substantive vs. perceptible information} \label{Section:Information}
Any physical realization of a computing device in the universe is evidently subject to the fundamental laws of nature. Hence we need to look into how these laws are influential in information processing. In any physical system, information is represented by different states of the system. For example, in many digital computers, different voltage levels are used as different states of a wire to represent different bits of information. Even in physical systems that are not for computing, e.g., a ball bouncing on a surface or a sugar cube dissolving in a glass of water, we still have different system states that bear information, even though we might not be interested in that information. From information theory it is known that, assuming the system state is denoted by the variable $X$, the amount of information represented by the system is \cite{Cover}:
\begin{equation}\label{eq:Shannon} 
H(X)=\mathbb{E}[\log_2 f_X(X)] 
\end{equation}
where $\mathbb{E}[\cdot]$ denotes expected value and $f_X(\cdot)$ denotes probability distribution function of $X$.

The branch of theoretical physics that mainly involves states of physical systems is statistical mechanics \cite{Susskind,Pathria} where states are considered at two different scales viz. microscopic and macroscopic scales. The microscopic state of a system, a.k.a., microstate, specifies the state of the system at the particle level. For example, it may be defined by specifying the positions and momenta of all the system particles individually \cite{Pathria}. The microstate of a system cannot be measured and known as it is impossible to individually track each particle of the system. On the other hand, the macroscopic state of a system, a.k.a., macrostate, involves the bulk properties of the system that are measureable such as temperature or volume \cite{Rex}. 
Based on the conventional concepts of microstate and macrostate, here we provide definitions for two concepts, called perfect microstate and perfect macrostate. Despite providing new definitions, the two defined concepts are not basically different from the conventional ones and are consistent with them. The definitions are just provided for the sake of disambiguation and to be specific and accurate enough to help develop the axiomatic theory. In the following, while providing the new definitions, it is discussed how they relate to the conventional ones.

\begin{definition1} [Perfect Microstate] \label{def:micro}
Let $A$ be a physical realization of a system. The \textbf{perfect microstate} of $A$, denoted by $\mathfrak{M}_A(t)$, represents the complete information in full detail about what the system is in reality at time $t$.
\end{definition1}

In statistical mechanics, although a microstate provides particle level information, it is sometimes defined with some lack of information. For example, one common way to specify microstates is to divide the phase space into cells and then specify which particle lies within which cell \cite{Rex}. Such a specification of microstates does not provide complete information about the system as when it specifies that a particle lies in a cell, it does not specify the exact location of the particle inside the cell. Indeed, a perfect microstate is a special case of a conventional microstate wherein the system is completely specified without any information loss. It is noteworthy that for a physical system $A$, the information represented by its perfect microstate $\mathfrak{M}_A(t)$ cannot be perceived completely as we know from statistical mechanics that the exact microstate of a system is impossible to be measured or known \cite{Rex}. Nevertheless, here we do not involve whether this information is perceptible or not. Regardless of the perceptibility, the system perfect microstate bears this information whose amount is $H(\mathfrak{M}_A(t))$ and throughout this paper we call this information the substantive information.

\begin{definition2} [Perfect Macrostate] \label{def:macro}
Let $A$ be a physical realization of a system. The \textbf{perfect macrostate} of $A$, denoted by $M_A(t)$, represents all the perceptible information about the system at time $t$.
\end{definition2}

It should be noted that, by perceptible information, we mean the information that can be perceived by an information-processing being (such as computers, measuring instruments, humans, etc.). In statistical mechanics, although the macrostate of a system provides some perceptible information about the system, the system may still bear some other perceptible information not represented by its macrostate. For example, in thermodynamics, the macrostate of a confined gas is commonly given by the tuple $(N,V,E)$ where $N$ is the number of gas particles, $V$ is the gas volume, and $E$ is the total energy of the gas \cite{Pathria}. Indeed, many other perceptible parameters such as gas temperature $T$ or pressure $P$ can be obtained from the tuple $(N,V,E)$. But there might be other perceptible information which are not given by $(N,V,E)$, e.g., the color of the gas or its chemical properties. Perfect macrostate is a specific case of conventional macrostate that contains all the perceptible information of the system and hence throughout this paper we call this information, whose amount is $H(M_A(t))$, the perceptible information.
By comparing Definitions \ref{def:micro} and \ref{def:macro} we can conclude that if we could perceive all the information of a real physical system then there would be no distinction between the system perfect microstate and perfect macrostate. But as mentioned previously, statistical mechanics implies that it is impossible to perceive the perfect microstate of a physical system and hence the distinction between perfect microstates and perfect macrostates is substantial.

Statistical mechanics mainly involves statistical analysis of microstates to see how the microscopic behavior of a system can be scaled up to explain its macroscopic behavior \cite{Susskind,Pathria,Rex}. However, in this paper we aim to have a different viewpoint into macrostates and microstates, one which is more appropriate for information processing. Indeed, we are interested to know the relationship between the information that the system perfect microstate bears, i.e., the substantive information $H(\mathfrak{M}_A(t))$ (regardless of the impossibility of perceiving this information) and the information that the system perfect macrostate bears, i.e., the perceptible information $H(M_A(t))$ (regardless of the fact that part of or all of it might not be important for whom who is observing the system macrostate). Analyzing this relationship between $H(\mathfrak{M}_A(t))$ and $H(M_A(t))$ has never been part of the classical statistical mechanics and even those few studies, e.g., \cite{Mezard,Jayness}, that consider both information theory and statistical mechanics. The rest of this section as well as the next two sections are devoted to provide mathematical equations for $H(\mathfrak{M}_A(t))$ and $H(M_A(t))$ and the relationship between them. 

We begin the discussion with the law of information conservation as one of the laws of nature. This law is indeed a direct conclusion of Liouville's theorem and holds in both classical and quantum mechanics \cite[Lecture 1]{Susskind}. While there are proofs for it in theoretical physics, we consider it as an axiom here to be used to prove other properties. This law indicates that the time evolution of any closed physical system is in such a way that the current state of the system uniquely determines its future and past states. It should be noted that the term \textquoteleft state\textquoteright \ in expressing the law of information conservation refers to what the system really is in the physical universe and not to what is perceived or measured by an observer (information-processing being). Therefore, the law of information conservation is indeed about perfect microstates. Here we formulate the law of information conservation as one of the axioms of the theory using the concept of perfect microstate defined in Definition \ref{def:micro}.

\begin{axinf} [The law of information conservation] \label{ax:inf}
Let $A$ be a closed physical system and $t_1$ and $t_2$ be two arbitrary time instants. Then $\mathfrak{M}_A(t_1)$, i.e., the perfect microstate of $A$ at time $t_1$, uniquely determines $\mathfrak{M}_A(t_2)$.
\end{axinf}

Using the notations of information theory \cite{Cover}, this axiom can be given in the form of mathematical equations. For this purpose we use the following lemma.

\begin{lmfunc} [Information theoretical notion of functions] \label{lm:func}
Variable $X$ uniquely determines the value of variable $Y$ (or in other words $Y$ is a function of $X$) if and only if the conditional entropy $H(Y|X)$ is equal to zero.
\end{lmfunc}

We have provided a proof for this lemma in Appendix 1. However, here we provide an interpretation for Lemma \ref{lm:func} to make the discussion more clear and easy to follow. In information theory, the conditional entropy $H(Y|X)$ indicates how much additional uncertainty the variable $Y$ introduces (or how much additional information it gives) assuming that we know the value of $X$. However, if the value of $X$ determines the value of $Y$ (or in other words, if $Y$ is a function of $X$), then by assuming that the value of $X$ is known, the variable $Y$ cannot introduce additional uncertainty (cannot provide additional information). This means that when we say $X$ determines the value of $Y$ ($Y$ is a function of $X$), it is equivalent to $H(Y|X)=0$. As mentioned in Appendix 1, this can be formally written as:
\begin{equation}\label{eq:function} 
H(Y|X)=0 \iff \exists f: Y=f(X) 
\end{equation}
Using Lemma \ref{lm:func}, we can formulate the law of information conservation (Axiom \ref{ax:inf}) as follows:
\begin{equation}\label{eq:infconvfull} 
\forall t_1,t_2: H(\mathfrak{M}_A(t_2)|\mathfrak{M}_A(t_1))=0 
\end{equation}

To give an insight into Eq.\ \ref{eq:infconvfull}, consider the two possible cases: $t_2>t_1$ and $t_2<t_1$. Without loss of generality, let us call $t_1$ the current time. When $t_2>t_1$, Eq.\ \ref{eq:infconvfull} indicates that the current perfect microstate of a closed system uniquely determines its future perfect microstates. We call this the property of \textbf{substantive determinism}. When $t_2<t_1$, Eq.\ \ref{eq:infconvfull} indicates that the current perfect microstate of a closed system uniquely determines its past perfect microstates or in other words, the perfect microstate of a closed system always holds the complete information about which microstates the system went through. We call this the property of \textbf{conserving substantive information}. In short, we can see that the law of information conservation has two time-symmetric parts that are the properties of substantive determinism and conserving substantive information. Based on Axiom \ref{ax:inf} (i.e., Eq.\ \ref{eq:infconvfull}) we have proved the following theorem in Appendix 2.

\begin{thmconst} [Constant amount of substantive information] \label{thm:const}
Let $A$ be a closed physical system. Then the amount of information represented by the system perfect microstate remains constant with time, i.e., $H(\mathfrak{M}_A(t))=Constant$.
\end{thmconst}

It must be emphasized that Theorem \ref{thm:const} should not be confused with the law of information conservation (Axiom \ref{ax:inf}) itself. Indeed, it is noteworthy that how the law of information conservation differs from other common conservation laws, e.g., energy conservation. Unlike the law of energy conservation which states that for any closed system, $E(t)=Constant$ (where $E(t)$ is the energy of the system at time $t$), the law of information conservation is not simply $H(\mathfrak{M}_A(t))=Constant$ and this equation is just one of the conclusions from the law of information conservation. The law is indeed $\forall t_1,t_2: H(\mathfrak{M}_A(t_2)|\mathfrak{M}_A(t_1))=0$ that indicates, for a closed system A, any perfect microstate determines all its past and future perfect microstates. It is also noteworthy that how the information theoretical concept of \textquoteleft conditional entropy\textquoteright \ is used to correctly formulate the law of information conservation mathematically.


Now let us consider another axiom. In statistical mechanics, while for each possible macrostate there are many possible microstates, which correspond to that single macrostate, each microstate corresponds to only one macrostate \cite{Pathria,Rex}. This implies that:

\begin{axmic} [Macrostate Determination by Microstate] \label{ax:mic}
For any physical system A, at any time $t$, the perfect microstate $\mathfrak{M}_A(t)$ uniquely determines the perfect macrostate $M_A(t)$.
\end{axmic}

It is noteworthy that this axiom does not necessarily require the system being closed. It can also be intuitively known from the definition of perfect microstate and perfect macrostate (Definitions \ref{def:micro} and \ref{def:macro}), that for any system (regardless of being closed or not) as the perfect microstate contains all the information about the system, it must determine all the information perceived from the system. 

Using Lemma \ref{lm:func}, we can formulate Axiom \ref{ax:mic} as follows:
\begin{equation}\label{eq:micmac} 
\forall t: H(M_A(t)|\mathfrak{M}_A(t))=0 
\end{equation}

But what about $H(\mathfrak{M}_A(t)|M_A(t))$? We know from statistical mechanics that for each macrostate $m_i$ there is a set of microstates, say $\Gamma(m_i)$, that correspond to that single macrostate \cite{Rex}. Therefore, it can be concluded that $M_A(t)$ does not uniquely determine $\mathfrak{M}_A(t)$ and hence $H(\mathfrak{M}_A(t)|M_A(t))\neq 0$. Indeed, in this paper we prove the following theorem, implying that $H(\mathfrak{M}_A(t)|M_A(t))$ gives the thermodynamic entropy, which is a result of central importance in the theory.

\begin{thmentropy} [Information theoretical notion of thermodynamic entropy] \label{thm:ent}
Let $A$ be a physical system. Within the axiomatic framework where thermodynamic entropy is defined, we have:
\begin{equation}\label{eq:thment} 
H(\mathfrak{M}_A(t)|M_A(t))=\frac{1}{k_B \ln{2}}\bar{S}_A(t)
\end{equation}

\noindent where $k_B$ is Boltzmann\textquoteright s constant and $\bar{S}_A(t)$ is the thermodynamic entropy of system $A$ at time $t$.
\end{thmentropy}

A proof for this theorem has been provided in Appendix 2. Indeed, since the inception of information theory, the resemblance between information theoretical entropy (Shannon entropy) and statistical notion of thermodynamic entropy (Boltzmann or Gibbs entropy) has implied that there are important ties between the two disciplines and even, many have believed that these two entropies are the same. Here, in this paper, it is not intended to totally dismiss that belief, but it must be emphasized that the proper and accurate information theoretical notion of thermodynamic entropy is in the form of conditional entropy as given by Theorem~\ref{thm:ent}.

One important point about thermodynamic entropy is that it is classically defined under the assumption of thermal equilibrium. Indeed, the literature in the context of thermodynamics and statistical mechanics commonly confined themselves to an axiomatic framework where only equilibrium states are considered (see for example \cite{Pathria}). Therefore, as the thermodynamic entropy has not been defined outside this framework, Theorem~\ref{thm:ent} must be inevitably restricted to this framework, too, as it is explicitly mentioned in the statement of the theorem. Nevertheless, while the traditional definitions of thermodynamic entropy, i.e., the original Clausius\textquoteright s definition \cite{Rex}, and Boltzmann\textquoteright s/Gibbs\textquoteright s definition \cite{Pathria}, depend on assumptions such as thermal equilibrium, the term $H(\mathfrak{M}_A(t)|M_A(t))$ depends only on the definitions of the microstate $\mathfrak{M}_A(t)$ and the macrostate $M_A(t)$ and does not require more restrictive assumptions like thermal equilibrium. Therefore, $H(\mathfrak{M}_A(t)|M_A(t))$ can be considered as a generalization for the concept of thermodynamic entropy. In short, we propose the following postulate: 

\begin{pstmain} [Generalization of Theorem~\ref{thm:ent}] \label{pst:main}
For any physical system A, the conditional entropy $H(\mathfrak{M}_A(t)|M_A(t))$ gives the thermodynamic entropy of the system at time $t$ without any precondition.
\end{pstmain}

Indeed, Theorem~\ref{thm:ent} can be considered as a strong evidence for Postulate~\ref{pst:main} as it indicates that if we limited ourselves only to systems in thermal equilibrium (just like what is done in classical thermodynamics and statistical mechanics  \cite{Pathria}), we would not need to consider Postulate~\ref{pst:main} as a postulate, since it would be a provable theorem. But the evidence is not just limited to Theorem~\ref{thm:ent}. Another important evidence for Postulate~\ref{pst:main} that we discuss in Section~\ref{Section:Conforming} is that we prove $H(\mathfrak{M}_A(t)|M_A(t))$ is additive without requiring to be confined in the axiomatic framework of classical thermodynamics and statistical mechanics. Additivity is an important property of thermodynamic entropy, which means if a physical system $A$ is made up of components $B$ and $C$, then we have: $\bar{S}_A(t)=\bar{S}_B(t)+\bar{S}_C(t)$. This property has been traditionally proved in the axiomatic framework of thermodynamics and statistical mechanics, e.g., with assumptions such as thermal equilibrium or equal a priori probabilities \cite{Pathria,Rex}. But in Section~\ref{Section:Conforming}, we will prove this property for $H(\mathfrak{M}_A(t)|M_A(t))$ without requiring such assumptions. In short, Postulate~\ref{pst:main} backed with the evidences provided by Theorem~\ref{thm:ent} and Theorem~\ref{thm:additivity} (in Section~\ref{Section:Conforming}) propounds $H(\mathfrak{M}_A(t)|M_A(t))$ as thermodynamic entropy in a general sense.

We have shown in Appendix 2 that based on Postulate \ref{pst:main}, the following theorem can be proved: 

\begin{thmMicroMacro} [Relationship between thermodynamic entropy and perceptible information] \label{thm:micmac}
Let $A$ be a closed physical system. Then at any time instant $t$, for the perceptible information $H(M_A(t))$ we have: 
\begin{equation}\label{eq:thmmicmac} 
C_{SI}=H(M_A(t))+\frac{1}{k_B \ln{2}}\bar{S}_A(t)
\end{equation}

\noindent where $C_{SI}$ is the substantive information of the closed system and is a constant value (see Theorem \ref{thm:const}), $k_B$ is Boltzmann\textquoteright s constant, and $\bar{S}_A(t)$ is the thermodynamic entropy of the system at time $t$.
\end{thmMicroMacro}

The importance of this theorem is that it shows how the perceptible information $H(M_A(t))$ and the thermodynamic entropy $\bar{S}_A(t)$ are related in a closed system. 

As mentioned in Section \ref{Section:Introduction}, in this paper we intend to develop a theory in the form of an axiomatic system to study the energy consumption of computing (information processing). To this end, so far in this section, we developed a basis including axioms, theorems, and a postulate involving substantive information $H(\mathfrak{M}_A(t))$, perceptible information $H(M_A(t))$, their relationship (e.g., $H(\mathfrak{M}_A(t)|M_A(t))$) and their relationship with thermodynamic entropy $\bar{S}_A(t)$. One question that may arise here is how these concepts relate to energy consumption. The answer is that energy consumption, at its core, is not mainly about energy, but rather is about thermodynamic entropy. Indeed, one of the fundamental laws of the universe is that energy is conserved (forget about the mass-energy conversion, which is irrelevant to our discussion). This means that energy cannot be really consumed and what we usually interpret as energy consumption is nothing but the increase in thermodynamic entropy. When thermodynamic entropy has to increase by a value $\Delta \bar{S}$, this usually happens by releasing the heat $\Delta \bar{S} \cdot T$ to an environment with the Kelvin temperature $T$ \cite{Pathria, Rex}, which is called consuming energy and this shows how energy consumption is indeed an interpretation of increase in thermodynamic entropy. Therefore, if one wants to have a profound theory about the energy consumption of information processing, one needs to address thermodynamic entropy as we have done in this section. In the next section, we exploit the basis developed in this section to further develop the theory, especially to address the energy consumption of information processing.

\section{Energy consumption of information processing} \label{Section:Energy} 
We begin the discussion of this section with Landauer's principle \cite{Landauer}, which is presumably the most well-known attempt to describe the reason why information processing requires energy consumption. This principle states that the reason of energy consumption is information erasure and whenever we erase one bit of information we need to release the heat ${k_B T \ln{2}}$ into the environment, where $k_B$ is Boltzmann\textquoteright s constant, and $T$ is the ambient temperature (i.e., the temperature of the environment where information processing happens) in Kelvin. The original rationale behind this was not formal proofs (see \cite{Landauer}), but rather arguments based on the second law of thermodynamics \cite{Landauer,Bennett3}, so that since the inception of the principle, it has been controversial and subject to serious critiques \cite{Earman, Shenker, Norton, Norton2}.

Here, we adopt a formal approach and exploit the basis developed in Section \ref{Section:Information} to formulate Landauer's principle in the form of a mathematical equation. This helps to disambiguate the principle, specify the condition under which the principle holds, and give a clear insight into how the principle is in effect.

It should be noted that when Landauer's principle talks about information loss, this loss occurs at the macroscopic level, i.e., for a physical system $A$, this principle involves $M_A(t)$. This is because: \emph{i})~The context of Landauer's work in \cite{Landauer} is about computers and information-processing machines. Clearly, any computer is intended to perform computations on perceptible information (otherwise the user cannot perceive and use the information) and based on the discussion in Section \ref{Section:Information}, the perceptible information of a system involves its macrostate (or more accurately, its perfect macrostate). \emph{ii})~As discussed in Section \ref{Section:Information}, we know from the law of information conservation (Axiom \ref{ax:inf}) that at the microscopic level the system information can never be lost\footnote{ Of course the law of information conservation is about closed systems, but even for systems that are not closed, they can still be considered as closed if we include their environment and in this case, if the system loses information at the microscopic level, this means that the information is moved from the system to its environment.} , which means that Landauer's principle (which basically involves information loss) must not be about substantive information, i.e., the information at the microscopic level.

Bearing in mind that Landauer's principle involves information at the macroscopic level, i.e., perceptible information, consider a physical system $A$, and let us see what the conditional entropy $H(M_A(t_2)$ $|M_A(t_1))$ means, where $t_1$ and $t_2$ are two arbitrary time instants. It is noteworthy to recall that we have already studied the microscopic counterpart, i.e., $H(\mathfrak{M}_A(t_2)|\mathfrak{M}_A(t_1))$, in Section \ref{Section:Information} and here we perform a similar analysis. Without loss of generality, let $t_1$ be the current time. When $t_2>t_1$, $H(M_A(t_2)|M_A(t_1))$ denotes how much additional uncertainty the future perfect macrostate $M_A(t_2)$ introduces, assuming that we know the current perfect macrostate $M_A(t_1)$. If the current perfect macrostate of a system cannot certainly determine its future perfect macrostates, this clearly means that the system does not behave in a deterministic manner and shows random behavior. Indeed, when $t_2>t_1$, $H(M_A(t_2)|M_A(t_1))$ is a measure of non-determinism (randomness) at the macroscopic level.

We can put this into the following definition: 
\begin{definition2.5} [Macroscopic Non-determinism (Randomness)] \label{def:macrand}
Let $A$ be a physical realization of a system and $t_1$ and $t_2$ be two arbitrary time instants such that $t_2>t_1$. Then $H(M_A(t_2)|M_A(t_1))$ is a measure of non-determinism (randomness) in the behavior of system $A$ at the macroscopic level as time goes from $t_1$ to $t_2$. 
\end{definition2.5}

One especially important case is when $t_2>t_1$ and $H(M_A(t_2)|M_A(t_1))=0$, which indicates that the current perfect macrostate of the system uniquely determines its future perfect macrostate at time $t_2$. We put this case into the following definition: 

\begin{definition3} [Macroscopic Determinism] \label{def:macdet} 
Let $A$ be a physical realization of a system. We say that the system is macroscopically deterministic if and only if\ \ $\forall t_1,t_2:\ t_2>t_1 \Rightarrow H(M_A(t_2)|M_A(t_1))=0$. 
\end{definition3}

When $t_2<t_1$, $H(M_A(t_2)|M_A(t_1))$ denotes how much additional uncertainty (additional information) the past perfect macrostate $M_A(t_2)$ introduces assuming that we know the current perfect macrostate $M_A(t_1)$. If the current perfect macrostate of a system cannot certainly determine its past perfect macrostates, this clearly means that the system has lost the information about its past macrostates. Indeed, when $t_2<t_1$, $H(M_A(t_2)|M_A(t_1))$ is a measure of information loss at the macroscopic level. We can put this into the following definition: 

\begin{definition4} [Macroscopic Information Loss] \label{def:macloss} 
Let $A$ be a physical realization of a system and $t_1$ and $t_2$ be two arbitrary time instants such that $t_2<t_1$. Then $H(M_A(t_2)|M_A(t_1))$ is the amount of information lost by system $A$ at the macroscopic level as time goes from $t_2$ to $t_1$. 
\end{definition4}

Therefore, when for two time instants $t_1$ and $t_2$, we have $t_2<t_1$ and $H(M_A(t_2)|M_A(t_1))=0$, we say that the system possesses reversibility in the time period from $t_2$ to $t_1$, since its state at the end of the period, i.e., $M_A(t_1)$, can be used to uniquely determine its state at the beginning of the period, i.e., $M_A(t_2)$. In short, we can provide the following definition: 

\begin{definition5} [Macroscopic Reversibility] \label{def:reversibility} 
Let $A$ be a physical realization of a system. We say that the system is macroscopically reversible if and only if\ \ $\forall t_1,t_2:\ t_2<t_1 \Rightarrow H(M_A(t_2)|M_A(t_1))=0$.
\end{definition5}

We now provide the following theorem that gives an equation helpful to shed insight into Landauer's principle. It should be noted that this theorem could be provided and proved before providing Definitions \ref{def:macrand} $\sim$ \ref{def:reversibility}, as this theorem does not depend on them. However, these definitions help us to interpret the theorem and conceive the meaning of its equation. We have proved this theorem in Appendix 2 where we have used Theorem \ref{thm:micmac} to provide the proof. 

\begin{thmLandauer} [The relationship between thermodynamic entropy, reversibility and determinism] \label{thm:landauer} 
Let $A$ be a closed physical system, and let $t_1$ and $t_2$ be two arbitrary time instants. Then we have: 
\begin{equation}\label{eq:landauer} 
\Delta \bar{S}=\bar{S}(t_2)-\bar{S}(t_1)=k_B \ln{2} \cdot [H(M_A(t_1)|M_A(t_2))-H(M_A(t_2)|M_A(t_1))]
\end{equation}

\noindent where $k_B$ is Boltzmann\textquoteright s constant, and $\bar{S}_A(t)$ is the thermodynamic entropy of the system at time $t$.
\end{thmLandauer} 

Considering the chain rule in information theory, which states that $H(X,Y)=H(X)+H(Y|X)$ holds for any pair of variables $X$ and $Y$\cite{Cover}, we can rewrite Eq.\ \ref{eq:landauer} of Theorem \ref{thm:landauer} in a simpler form as stated by Corollary \ref{cor:simpler} \footnote{Indeed, this corollary has been already proved within the proof of Theorem~\ref{thm:landauer}.}.

\begin{corSimpler} [A simpler form of Theorem~\ref{thm:landauer}] \label{cor:simpler} 

\emph{Eq.~\ref{eq:landauer}} can also be written as:
\begin{equation}\label{eq:simplerform} 
\Delta \bar{S}=\bar{S}(t_2)-\bar{S}(t_1)=k_B \ln{2} \cdot [H(M_A(t_1))-H(M_A(t_2))]
\end{equation}
\end{corSimpler}

Although Eq.~\ref{eq:simplerform} looks simpler than Eq.~\ref{eq:landauer}, but Eq.~\ref{eq:landauer} might be more preferable in order to shed insight into the concept of thermodynamic entropy as it shows how thermodynamic entropy relates to reversibility and non-determinism (randomness). It is noteworthy that Eq.\ \ref{eq:landauer} of Theorem \ref{thm:landauer} (and also Eq.~\ref{eq:simplerform} of Corollary \ref{cor:simpler}) is time-symmetric in the sense that if we interchange $t_1$ and $t_2$ we will still have the same equation. Therefore, it makes no difference for Theorem \ref{thm:landauer} whether to have $t_2>t_1$ or $t_2<t_1$. Hence, let us assume $t_2>t_1$, and consequently, as mentioned in Definitions \ref{def:macrand} and \ref{def:macloss}, $H(M_A(t_2)|M_A(t_1))$ is the measure of macroscopic non-determinism and $H(M_A(t_1)|M_A(t_2))$ is the measure of macroscopic information loss. Therefore, Eq.\ \ref{eq:landauer} of Theorem \ref{thm:landauer} shows how the variation of thermodynamic entropy with time relates to macroscopic determinism and macroscopic reversibility. One specific and important case is when a closed physical system behaves in a completely deterministic manner at the macroscopic level. In this case, as mentioned in Definition \ref{def:macdet}, for all values of $t_1$ and $t_2$ that $t_2>t_1$, we have $H(M_A(t_2)|M_A(t_1))=0$ and hence we can rewrite Theorem \ref{thm:landauer} as follows:

\begin{corLandauer} [Landauer's Principle] \label{cor:landauer}
Let $A$ be a closed physical system that behaves in a completely deterministic manner at the macroscopic level, and let $t_1$ and $t_2$ be two arbitrary time instants such that $t_2>t_1$. Then we have:
\begin{equation} \label{eq:reallandauer} 
\Delta \bar{S}=\bar{S}(t_2)-\bar{S}(t_1)=k_B \ln{2} \cdot H(M_A(t_1)|M_A(t_2))
\end{equation}

\noindent where $k_B$ is Boltzmann\textquoteright s Constant, and $\bar{S}_A(t)$ is the thermodynamic entropy of the system at time $t$.
\end{corLandauer}

It can be seen that Eq.\ \ref{eq:reallandauer} is what known as Landauer's principle in the form of a mathematical equation. This equation indicates that for every bit of the lost information $H(M_A(t_1)|M_A(t_2))$ the thermodynamic entropy increases by $k_B \ln{2}$. It should be noted that in the context of thermodynamics, a thermodynamic entropy increase of $\Delta \bar{S}$ usually occurs by releasing the heat $\Delta \bar{S} \cdot T$ into a heat bath (environment) with the Kelvin temperature $T$, and this is why Landauer's principle is usually phrased in the form of releasing the heat $k_B T\ln{2}$ rather than increasing thermodynamic entropy by $k_B \ln{2}$.

Aside from formulating Landauer's principle in the form of a mathematical equation, the important point here is that Landauer's principle is only valid for the systems with completely deterministic behavior at the macroscopic level. Indeed, it can be concluded from Eq.\ \ref{eq:landauer} that if the condition of macroscopic determinism (Definition \ref{def:macdet}) does not hold (i.e., $H(M_A(t_2)|M_A(t_1))\neq 0$ for $t_2>t_1$), Landauer's principle will be violated. This is because in the case of having lost information (i.e., $H(M_A(t_1)|M_A(t_2))\neq 0$), information loss and non-determinism (i.e., $M_A(t_2)|M_A(t_1))$ and $M_A(t_1)|M_A(t_2))$) can cancel each other in Eq.\ \ref{eq:landauer}, so that we will have no increase in thermodynamic entropy despite losing information. The author believes that the assumption of macroscopic determinism (Definition \ref{def:macdet}), which is crucial for the validity of Landauer's principle, has been overlooked in previous research and probably this vagueness has been one of the main sources of notable controversy on the principle.

Indeed, macroscopic determinism is an implicit assumption for many people, especially for engineers, as something of causality. For example, people usually expect that a well-designed system, say system $A$, is no place for unknown or undetermined behavior and must operate as designed.  It should be emphasized that this is indeed what people believe about what they perceive from the system, which is equal to the condition $H(M_A(t_2)|M_A(t_1))=0$. Even in the notable experimental work of \cite{Hong}, the author believes that macroscopic determinism is implicitly assumed, since the work in [1] has not considered or introduced any source for randomness (non-determinism) at the macroscopic (perceptible) level and this is why it reports in favor of Landauer's principle. However, we must note that based on the fundamental laws of the universe, strict causality (or determinism) only exists at the microscopic level (i.e., we surely have $H(\mathfrak{M}_A(t_2)|\mathfrak{M}_A(t_1))= 0$), but at the macroscopic level, the laws of the universe do not necessitate determinism (i.e., we might have $H(M_A(t_2)|M_A(t_1))\neq 0$) and hence macroscopic determinism, as defined in Definition \ref{def:macdet}, can be violated and we must expect random (non-deterministic) behavior at the macroscopic level for closed physical systems.

From this discussion, we can see that non-determinism (or randomness) at the macroscopic level must play an important role in energy consumption (increase in thermodynamic entropy) when performing information processing, just like the important role that information loss has. This is indeed a direct conclusion from Eq.\ \ref{eq:landauer} whose time symmetry puts both $H(M_A(t_1)|M_A(t_2))$ and $H(M_A(t_2)|M_A(t_1))$ (i.e., information loss and randomness) at the same level of importance. However, assuming macroscopic determinism (the implicit assumption of previous work) leaves no room for random behavior at the macroscopic level and hence reduces Eq.\ \ref{eq:landauer} to Eq.\ \ref{eq:reallandauer}.

At the end of this section, it is worth highlighting several important points: 

\emph{I}) If we assume macroscopic determinism, which means limiting ourselves to Eq.\ \ref{eq:reallandauer} (i.e., Landauer's principle), to reduce energy consumption (i.e., to ameliorate the increase in thermodynamic entropy) in computation, we must reduce information loss (i.e., to reduce $H(M_A(t_1)|M_A(t_2))$ in Eq.\ \ref{eq:reallandauer}). This justifies the use of reversible computing \cite{Frank} to reduce the energy consumption of computers, as it avoids information loss. Nevertheless, here in this paper we aim at considering the other side of the coin, which is non-determinism or randomness. In the next section we discuss how exploiting randomness in computation, besides reversibility, can let us go even beyond the energy saving that reversible computing offers. 

\emph{II}) It is noteworthy that when, in this paper, we address randomness, we mean real randomness where the current system macrostate (or more accurately, perfect macrostate) cannot be determined by its previous macrostates. Pseudo randomness, which is sometimes used in computation, is not the subject of the discussion here, since in pseudo randomness the system state is uniquely determined by its previous states and hence does not result in the non-determinism that we addressed here. 

\emph{III}) While the original rationale behind Landauer's principle comes from the second law of thermodynamics, here we have not assumed the second law while deducing the principle (i.e., Eq.\ \ref{eq:reallandauer}). Indeed, we have used the property of macroscopic determinism and the basis developed in Section \ref{Section:Information} which is mainly based on the law of information conservation. 

\emph{IV}) It is worth emphasizing that Eq.\ \ref{eq:landauer} (Theorem \ref{thm:landauer}) and Eq.\ \ref{eq:reallandauer} (Corollary \ref{cor:landauer}) are completely general in this sense that they are not just about information processing systems (computers). Rather they are based on fundamental laws of nature and hence hold for any physical system. For example, Eq.\ \ref{eq:reallandauer}  not only indicates that if we erase a bit of information from a computer memory, we will have an increase in thermodynamic entropy, but it is also so general that indicates, in any physical system, whenever thermodynamic entropy increases, this must be in the form of information loss at the macroscopic level. An example of this is when an object with the initial temperature $T_i$ comes into thermal contact with a heat bath of temperature $T_f$ and finally reaches thermal equilibrium with the temperature $T_f$. In thermodynamics, this is a classic example of increase in thermodynamic entropy, but here we look at it as a loss of perceptible information, where at least the initial temperature of the object has become imperceptible.

\section{Randomness: Beyond the reversible computing} \label{Section:Randomness} 
So far, theoretical studies of energy consumption in computation have been oriented towards reversible computing, which has roots in Landauer's principle. However, as mentioned in Section \ref{Section:Energy}, Landauer's principle, which is captured in Eq.\ \ref{eq:reallandauer} of Corollary \ref{cor:landauer}, is based on the implicit assumption of macroscopic determinism (defined in Definition \ref{def:macdet}) for a closed system. But it should be noted that there is no rule in nature that necessitates macroscopic determinism, and indeed macroscopic determinism is usually the way engineers look into their devised systems that must follow specific behaviors. Clearly, based on Eq.\ \ref{eq:landauer}, if we can find a way to violate macroscopic determinism and have $H(M_A(t_2)|M_A(t_1)) \neq 0$ when $t_2 > t_1$, we can alleviate the increase in thermodynamic entropy. In other words, in Eq.\ \ref{eq:landauer}, besides the term $H(M_A(t_1)|M_A(t_2))$ that denotes information loss, the term $H(M_A(t_2)|M_A(t_1))$ that denotes randomness (macroscopic non-determinism) can also provide a means to control energy consumption of computing. Our main intention in this section is to investigate how it is possible to exploit randomness along with reversibility (i.e., no information loss) to address the energy consumption of computing, as predicted by Eq.\ \ref{eq:landauer}.

There are already approaches in computation that use randomness for different purposes such as time or space efficiency \cite{Sipser}, but, as will be indicated by the discussion in this section, the issue of energy consumption (i.e., increase in thermodynamic entropy) is different and we cannot use the same approaches. Let us investigate how Eq.\ \ref{eq:landauer} gives clues to reduce energy consumption of computing. Consider a Turing machine (the most common mathematical model of computers \cite{Sipser}) TM and its surrounding environment that are in thermal equilibrium at temperature $T$ as a closed physical system $A$. Here, to avoid unnecessary complication, we suppose that the environment is as simple as possible, e.g., a heat bath with temperature $T$, as commonly assumed in the context of thermodynamics and statistical mechanics. Also, to avoid unnecessary complication, we suppose that all perceptible features of this system that are not relevant to computation, such as color, weight, and temperature are fixed and hence bear no information (it can be easily shown in information theory that a variable $X$ with an unchanging fixed value bears no information and $H(X)=0$). In this case, the following lemma, which has been proved in Appendix 1, can be used to exclude all these non-computational features from Eq.\ \ref{eq:landauer} and focus just on computing. 

\begin{lmremove} [Removing non-contributing variables] \label{lm:remove} 
Assuming that variables $W$ and $Y$ bear no information \emph{(}i.e., $H(W)=0$ and $H(Y)=0$\emph{)}, we have $H(W,X|Y,Z)=H(X|Z)$. 
\end{lmremove}

Now let $t_1$ be the time that TM begins its computation with an input string $I$ on its tape and $t_2$ be the time that TM finishes its computation and halts with the output string $O$ on its tape. Lemma~\ref{lm:remove} indicates that, for $H(M_A(t_1)|M_A(t_2))$ and $H(M_A(t_2)|M_A(t_1))$ in Eq.~\ref{eq:landauer}, we can exclude all parts of the system perfect macrostate, $M_A(t)$, that are not relevant to computation (as we assumed that they are fixed and hence bear no information) and retain only those parts that take part in computation. Doing so, we should just consider what is on the tape of TM, i.e., from Lemma \ref{lm:remove} we conclude: 
\begin{equation}\label{eq:Removing} 
H(M_A(t_1)|M_A(t_2))=H(I|O), \quad H(M_A(t_2)|M_A(t_1))=H(O|I)
\end{equation}

Hence, we can rewrite Theorem\ \ref{thm:landauer} (Eq.\ \ref{eq:landauer}) as follows:

\begin{corTheorem} [Variation of thermodynamic entropy in a Turing machine] \label{cor:Theorem} 
Suppose a Turing machine TM and its environment comprise a physical system $A$ and suppose all perceptible features of system $A$ that are not relevant to computation are fixed and hence bear no information. Let $t_1$ be the time that TM begins its computation with an input string $I$ on its tape and $t_2$ be the time that TM finishes its computation and halts with the output string $O$ on its tape. Then the variation of thermodynamic entropy in the Turing machine thermodynamic due to this computation is\footnote{ Eq.~\ref{eq:TM} can also be written as $\Delta \bar{S}=k_B \ln{2} \cdot [H(I)-H(O)]$ because of the chain rule (i.e., $H(X,Y)=H(X)+H(Y|X)$ \cite{Cover}). However, in this section we mainly use Eq.~\ref{eq:TM} to support our discussion. }: 
\begin{equation}\label{eq:TM} 
\Delta \bar{S}=\bar{S}(t_2)-\bar{S}(t_1)=k_B \ln{2} \cdot [H(I|O)-H(O|I)]
\end{equation}

\noindent where $k_B$ is Boltzmann\textquoteright s constant, and $\bar{S}_A(t)$ is the thermodynamic entropy of the system at time $t$. 
\end{corTheorem}

It should be noted that the part of the macrostate of system $A$ that involves computation is indeed what is called Turing machine configuration \cite{Sipser}, which consists of not only the tape contents but also the head position and the state of the control. However, TM can be devised so that the head position and the control state are the same for both the beginning and the end time (i.e., the time that TM begins the computation and the time it halts). Therefore, the head position and the control state can also be excluded, like non-computational features, and this is why we have only retained the tape contents in Eqs.~\ref{eq:Removing}, and \ref{eq:TM}. 

In the remainder of this section, we will see that Eq.~\ref{eq:TM} is very useful to calculate the minimum energy consumption of a computation, required by the laws of nature, but first we clarify several important points about this equation:

\emph{I}) While Eq.~\ref{eq:TM} is formulated in terms of tape contents of Turing machines, we should place emphasis that it was developed based on the laws of nature and mainly the law of information conservation, as it can be seen by following how the equation has been developed. 

\emph{II}) It should be emphasized that this equation indeed gives the minimum energy consumption of a computer because it is the energy that computing necessitates its consumption. The energy consumption of today's computers is much higher because they consume energy beyond what is necessitated by computing. This is due to violating the assumption where all the perceptible features of the system that are not relevant to computation are fixed and hence bear no information. Practically speaking, there are many perceptible features, e.g., currents in wires, charges in capacitors, temperature of components, etc., in today's computers that bear information which is useless and unnecessary for the computation performed by the computer. But based on Eq.~\ref{eq:landauer}, as the system carelessly loses such information, this will inevitably results in energy dissipation.

\emph{III}) As mentioned from the beginning of this paper, here we have adopted a pure theoretical approach to analyze the energy consumption of computing. Therefore, in this paper we do not involve implementation or technical issues. Especially, when we refer to the terms like physical implementation or physical laws, this does not mean we want to involve technical aspects; rather we want to indicate that relevant concepts from theoretical physics, that are essential to analyze the energy consumption of computing, have been considered and included. In this regard, while Eq.~\ref{eq:TM} provides a theoretical limit as the minimum energy consumption of a computation, required by the laws of nature, it does not directly provide a way to perform computing or to implement computers that have low energy consumption close to theoretical limits.

\emph{IV}) While, as mentioned in the previous item, Eq.~\ref{eq:TM} does not provide an implementation method and is just about theoretical limits, yet it provides some clues and guides to achieve low energy consumption in computing. Two important clues, as design guides, are reversibility and randomness. It is clear from Eq.~\ref{eq:TM} that to reduce $\Delta \bar{S}$ (thereby reducing energy consumption), we must make $H(I|O)$ as low as possible and $H(O|I)$ as high as possible. $H(I|O)$ denotes how much additional information the input string $I$ gives, provided that we already know the output string $O$. It is clear that if the complete information of the input $I$ is conserved in the output $O$, we will have $H(I|O)=0$, and this is indeed the basis for reversible computing, where we avoid forgetting the information provided by the input. On the other hand, $H(O|I)$ denotes how much additional information (uncertainty) the output string $O$ gives, provided that we already know the input string $I$. The common practice in computer science and engineering is that each input results in one definite output and hence $H(O|I)$ usually has the lowest possible value, i.e., $0$. To increase $H(O|I)$ (in order to reduce $\Delta \bar{S}$ and, thereby, energy consumption), the output $O$ must show more uncertainty despite having a specific input $I$. This means that we should include some source of randomness in the computation and retain the random information in the output to increases its uncertainty. While reversible computing has been considered in previous research studies on low energy computation, none of them has noticed randomness and its similar prominence.

\emph{V}) It can be seen from Eq.~\ref{eq:TM} that the energy consumption only depends on the initial and final states of the computation and does not depend on the way the computation progresses. It is interesting that the same holds in thermodynamics for devices such as heat engines and refrigerators, where the thermodynamic entropy is known to be a state variable, i.e., to find the entropy change $\Delta \bar{S}$, the path from the initial state to the final state in the state space is not important, and only the initial and final states really matter. 
\\~

Now, to give some insight into how Eq.~\ref{eq:TM} can be useful to calculate the lower theoretical limit of energy consumption of computing, we consider two examples. One is sorting an array of $n$ different objects and the other is whether an $n$-digit decimal number is divisible by~3 or not.

\noindent\hfil\rule{0.5\textwidth}{.4pt}\hfil

\textbf{Example 1}: How much is the lower theoretical limit of energy consumption required for sorting an array of $n$ different objects? 

To answer this question, we need to calculate $H(I|O)$ and $H(O|I)$. To calculate $H(I|O)$, we first consider the common case, i.e., when reversible computing is not exploited, which means that the output does not have to retain the input information. The conditional entropy $H(I|O)$ gives the amount of uncertainty (information) introduced by the input array, given that the output (sorted) array is known. By knowing the output array, the only source of uncertainty in the input could be the permutation of array elements. Since we have $n$ different objects, the number of permutations is $n!$. Assuming that each permutation in the input is equally probable, $H(I|O)$ is:
\begin{equation}\label{eq:exm1_1} 
H(I|O)= \log_2 n! 
\end{equation}

To calculate $H(O|I)$, we consider the common practice in computing where a computation is supposed to just give the answer to a given problem as the output, without providing any extra information. The conditional entropy $H(O|I)$ gives the amount of uncertainty (information) introduced by the output array, given that the input array is known. Obviously, if we know what an array is, then the result of sorting cannot show any uncertainty and it must be a completely definite output, i.e., the sorted array. Therefore, we have: 
\begin{equation}\label{eq:exm1_2} 
H(O|I)= 0 
\end{equation}

Now, if we substitute from Eqs.~\ref{eq:exm1_1} and \ref{eq:exm1_2} into Eq.~\ref{eq:TM}, we obtain: 
\begin{equation}\label{eq:Sort} 
\Delta \bar{S}=k_B \ln{2} \cdot (\log_2 n!-0)=k_B \ln{n!} \approx k_B (n\ln{n} - n) 
\end{equation}

\noindent where we have used Stirling\textquoteright s approximation, i.e., $\ln{n!} \approx n\ln{n} - n$. As a numerical example, suppose we want to sort an array with 1000 elements in the room temperature 25$^{\circ}$C. In this case, the temperature in Kelvin is $T=(25+273.15)=298.15\textrm{K}$, and since $k_B = 1.38 \times 10^{-23} \textrm{J}\cdot\textrm{K}^{-1}$, Eq.~\ref{eq:Sort} indicates that the theoretical minimum energy, which must be dissipated, is:
\begin{equation}\label{eq:Numerical} 
\Delta \bar{S}\cdot T \approx k_B (n\ln{n} - n) \cdot T = 1.38 \times 10^{-23} \times (1000\ln{1000}-1000) \times 298.15 = 2.43 \times 10^{-17} \textrm{J}
\end{equation}

Now suppose that we exploit reversible computing, i.e., the input information is retained in the output. One possible way to do this is that, besides the sorted array, the output must contain an intact copy of the input array. In this case, since the output $O$ contains the input $I$, knowing the output leaves no room for uncertainty in the input and we have $H(I|O)=0$. With respect to $H(O|I)$, just like the previous case, if we know what the input array is, the output cannot show any uncertainty as it contains the sorted array and a copy of the input. Therefore, we also have $H(O|I)=0$, and hence $\Delta \bar{S} = 0$. This shows the prominence of reversible computing as it means that if we exploit reversible computing, at least from a theoretical point of view, it is possible to sort an array without dissipating any energy. Of course, this is not a novel or astounding theoretical observation and it has been already mentioned in previous studies such as \cite{Feynman, Bennett, Bennett2}. But what is really important here and has not been addressed in previous studies, especially as compared to reversible computing, is that randomness can be exploited to achieve the same result. 

To see the impact of randomness, suppose that TM is equipped with some mechanism that can introduce uncertainty. This time, suppose that when TM gives the output, not only it must provide the sorted array, but also must provide a shuffled version of the input array generated by exploiting the randomness mechanism. Suppose that the shuffled array is generated in a way that any permutation of elements (out of $n!$ possible permutations) is equally probable. In this case, since the output contains only the sorted and shuffled arrays, the original order in the input array cannot be known from the output and hence we have $H(I|O)= \log_2 n!$. On the other hand, by knowing the input, that part of the output which is the sorted array cannot introduce any uncertainty, but the other part which is the shuffled array is a source of uncertainty and since all $n!$ permutations are equally probable, we have $H(O|I)= \log_2 n!$. This time, we have $\Delta \bar{S} = \log_2 n! - \log_2 n! = 0$ and again, like when we exploited reversible computing, we have $\Delta \bar{S} = 0$, which means that from a theoretical point of view it is possible to sort an array without dissipating any energy. But the notable point here is that this time we have not exploited reversibility (as the output does not retain the complete information of the input), rather we have exploited uncertainty (randomness) to reach the same energy saving. 

\noindent\hfil\rule{0.5\textwidth}{.4pt}\hfil

\textbf{Example 2}: How much is the theoretical minimum energy consumption required for deciding whether an $n$-digit decimal number is divisible by~3? 

Just like the previous example, we need to calculate $H(I|O)$ and $H(O|I)$. First we consider the case where reversible computing is not exploited. To calculate $H(I|O)$, we note that according to the definition of conditional entropy, we have \cite{Cover}:
\begin{equation}\label{eq:condent4DIV} 
H(I|O)=\underset{s}{\mathbb{E}}[H(I|O=s)]=\sum_{s\in D_O} [P(O=s)H(I|O=s)]
\end{equation}

\noindent where $D_O$ is the set of all possible values of $O$ and $H(I|O=s)$ is:
\begin{equation}\label{eq:condent4DIV2} 
H(I|O=s)=-\sum_{r\in D_I} [P(I=r|O=s)\log_2 P(I=r|O=s)]
\end{equation}
\noindent where $D_I$ is the set of all possible values of $I$.

In this problem we have $D_O=\{\textrm{Yes}, \textrm{No}\}$, since the input is either divisible by~3 or not. Therefore, Eq.~\ref{eq:condent4DIV2} can be rewritten as: 
\begin{equation}\label{eq:condent4DIVunfold} 
H(I|O)=P(O=\textrm{Yes})H(I|O=\textrm{Yes})+P(O=\textrm{No})H(I|O=\textrm{No})
\end{equation}

It can be easily shown that from all $n$-digit decimal numbers, exactly $\frac{10^n+2}{3}$ of them are divisible by~3. Therefore, assuming that all the numbers are equally probable to be the input, we conclude: 
\begin{equation}\label{eq:values} 
\begin{split}
&P(O=\textrm{Yes})=\frac{1}{3}+\frac{2}{3\times 10^n}, \quad
H(I|O=\textrm{Yes})=\log_2 \frac{10^n+2}{3}, \\
&P(O=\textrm{No})=\frac{2}{3}-\frac{2}{3\times 10^n}, \quad
H(I|O=\textrm{No})=\log_2 \frac{2\times 10^n-2}{3}
\end{split}
\end{equation}

If we substitute from Eq.~\ref{eq:values} to Eq.~\ref{eq:condent4DIVunfold}, we have:
\begin{equation}\label{eq:4DIVfinal} 
H(I|O)=[(\frac{1}{3}+\frac{2}{3\times 10^n}) \log_2 \frac{10^n+2}{3}]+[(\frac{2}{3}-\frac{2}{3\times 10^n})\log_2 \frac{2\times 10^n-2}{3}]
\end{equation}

To calculate $H(O|I)$, like the previous example, we first consider the common practice in computing where a computation is supposed to just give the answer to a given problem as the output, without providing any extra information. In this case, if we know what the input number is, the output cannot show any uncertainty because if the input is divisible by~3, the output is certainly \textquoteleft Yes\textquoteright, and otherwise the output is certainly \textquoteleft No\textquoteright. Therefore, we have $H(O|I)= 0$ and hence $\Delta \bar{S}=k_B \ln{2} \cdot H(I|O)$, where $H(I|O)$ is given by Eq.~\ref{eq:4DIVfinal}. 

Now suppose that we exploit reversible computing, i.e., the input information is retained in the output. One possible way to do this is that, besides a \textquoteleft Yes\textquoteright \ or \textquoteleft No\textquoteright \ as the answer, the output must contain an intact copy of the input number. In this case, since the output $O$ contains the input $I$, knowing the output leaves no room for uncertainty in the input and we have $H(I|O)=0$. With respect to $H(O|I)$, if we know what the input number is, the output cannot show any uncertainty as it contains the answer (\textquoteleft Yes\textquoteright \ or \textquoteleft No\textquoteright) and a copy of the input number. Therefore, we also have $H(O|I)=0$, and hence $\Delta \bar{S} = 0$. Like in the previous example, this means that if we exploit reversible computing, at least from a theoretical point of view it is possible to determine the divisibility by~3 without dissipating any energy.

As also mentioned in the previous example, this observation that reversibility can theoretically result in zero energy consumption is not novel. Nevertheless, it is notable that how the theory developed in this paper confirms and conforms to this property of reversibility. Furthermore, it is worth re-emphasizing here that what is really important and has not been addressed in previous work is that randomness can be exploited to achieve the same result as obtained by reversibility. This is what we see in the rest of the example. 

Suppose that the Turing machine TM is equipped with some mechanism that can introduce uncertainty. This time, suppose that when TM gives the output, not only it must provide a  \textquoteleft Yes\textquoteright \ or \textquoteleft No\textquoteright \ as the answer, but also must provide a random $n$-digit decimal number so that when the answer is \textquoteleft Yes\textquoteright, the random number is drawn uniformly from $n$-digit decimal numbers that are divisible by~3, and if the answer is \textquoteleft No\textquoteright, the random number is drawn uniformly from $n$-digit decimal numbers that are not divisible by~3. In this case, since the output contains only a \textquoteleft Yes\textquoteright \ or \textquoteleft No\textquoteright \ as well as a random $n$-digit number, the input number cannot be known from the output and the only thing that we can know from the output is the divisibility of the input by~3. Hence, $H(I|O)$ is still given by Eq.~\ref{eq:4DIVfinal}. On the other hand, by knowing the input, that part of the output which is either \textquoteleft Yes\textquoteright \ or \textquoteleft No\textquoteright \ cannot introduce any uncertainty, but the other part which is the random $n$-digit number is a source of uncertainty and it can be shown that in this case $H(O|I)$ is equal to $H(I|O)$ given by Eq.~\ref{eq:4DIVfinal}. Therefore, we again have $\Delta \bar{S} = 0$, which is like when we used reversible computing. Like in the previous example, the notable point here is that this time we have not exploited reversibility (as the output does not retain the complete information of the input), rather we have exploited uncertainty (randomness) to reach the same energy saving. 

\noindent\hfil\rule{0.5\textwidth}{.4pt}\hfil
\\~

There are several important points about these two examples that must be highlighted:
\begin{itemize}
\item We have seen that randomness can be exploited, like reversibility, to achieve low energy consumption. The important point is that in both the cases, TM should provide some information at the output which is not required by the problem and hence is useless from the viewpoint of the user. For instance, in the first example, not only TM provides a sorted array at the output, which is required by the problem, but also it either provides a copy of the unsorted input array at the output (when reversibility is exploited), or provides a random permutation of the input array at the output (when randomness is exploited). Such extra information at the output is not required for solving the problem, but producing and keeping them is needed to avoid energy dissipation; otherwise, if we throw away this extra information (i.e., erase or forget the extra information at the macroscopic level), energy will be dissipated as required by Eq.~\ref{eq:TM}.

\item It can be seen that to evaluate the minimum energy consumption of a computation, we do not need to involve implementation issues such as the algorithm which is used to perform the computation or the hardware which is used to implement the computer. This is because, based on Eq.~\ref{eq:TM}, what we need to calculate the minimum energy consumption of a computation is the information theoretic terms $H(I|O)$ and $H(O|I)$ that involve the information that input and output bear, regardless of the algorithm and the hardware of the system. 

\item In these two examples, we have not considered the simultaneous use of reversibility and randomness to avoid complication. But this simultaneous use and its important consequences will be discussed in the next three sections.

\end{itemize}

In the next section, we will have a look at exploiting macroscopic non-determinism (randomness) to achieve low energy consumption in computing. We will discuss that for some important classes of problems and applications, macroscopic non-determinism can be even more effective as compared to reversibility.

\section{Randomness in solving NP-complete problems} \label{Section:NP} 
One important point in the previous section is that in both the case of exploiting reversibility and the case of exploiting randomness for low energy consumption, we ended up with some extra information at the output that despite being useless from the viewpoint of the user, must be saved and cannot be erased; otherwise, leads to an increase in thermodynamic entropy (energy dissipation) based on Eq.~\ref{eq:TM}. This problem can get worse since computers are usually used repeatedly to process different inputs one after another. In this case, our low-dissipating computer has to keep some information for every input, which means it will soon run out of memory and will be forced to erase the information, which leads to energy dissipation and makes the exploitation of reversibility or randomness ultimately useless. Here in this section we provide an approach to tackle this problem with the use of randomness. The approach is provided for NP-complete problems, although we will discuss that it can also be exploited for many other problems. The reasons for considering NP-complete problems are: \emph{i}) We will argue that the proposed approach is well-suited for this class of problems, and \emph{ii}) NP-complete problems appear in various contexts with important practical applications \cite{Sipser}. This means that the proposed approach has a high potential to be practically useful.

The NP-complete problem that we consider here is the satisfiability problem and, from the theory of NP-completeness, we know that what we discuss here can be generalized to all other NP-complete problems. The satisfiability problem is: \textquotedblleft For a given Boolean expression, is there any assignment of values to the expression variables that can satisfy the expression (i.e., turn it to a true statement)?\textquotedblright \ Some may just want to know if such an assignment exists, but do not need to know the satisfying assignment. Here, we consider the problem where the satisfying assignment is also required. Suppose that we have a computer that solves this problem so that when we give it a Boolean expression as input, it provides a satisfying variable assignment as output. For example, suppose that we have the following Boolean expression:
\begin{equation} \label{eq:CNF} 
F=(p\lor \lnot r)\land (\lnot q\lor \lnot r)\land (\lnot p \lor \lnot q \lor s)\land (t\lor \lnot u\lor r)\land (r\lor \lnot v)
\end{equation}

If we input this expression to the computer, one possible output is the assignment $(p,q,r,s,t,u,v)=(\textrm{T},\textrm{F},\textrm{F},\textrm{F},\textrm{F},\textrm{F},\textrm{F})$, where F and T denote \textquoteleft False\textquoteright ~and \textquoteleft True\textquoteright ~respectively. However, there are also other assignments that can satisfy the same expression, such as $(p,q,r,s,t,u,v)=(\textrm{T},\textrm{F},\textrm{F},\textrm{T},\textrm{F},\textrm{F},\textrm{F})$ or $(p,q,r,s,t,u,v)=(\textrm{F},\textrm{F},\textrm{F},\textrm{F},\textrm{F},\textrm{F},\textrm{F})$. Indeed, as it can be seen from this example, one characteristic of NP-complete problems, which is important for the discussion here, is that they usually do not have a unique solution and there are many possible solutions for them. If the computer that solves the satisfiability problem adopts a deterministic approach, for each Boolean expression, it will always provide a fixed and specific solution among the possible solutions. However, here we want to have a computer that exploits a randomness mechanism to achieve macroscopic non-determinism, so that for any Boolean expression, it gives equal probability to all satisfying assignments to be the output solution.

Now let us analyze how we can ameliorate the increase in thermodynamic entropy (energy dissipation) by the use of this approach. It is worth emphasizing that in this analysis we do not intend to use reversibility, as we just want to focus on and study the impact of randomness. Therefore, we assume that the computer only provides the satisfying assignment as the output without requiring making it fully reversible by, for example, copying the input to the output. For this analysis, we use Eq.~\ref{eq:TM}, i.e., $\Delta \bar{S}=k_B \ln{2} \cdot [H(I|O)-H(O|I)]$. As seen in Section~\ref{Section:Randomness}, to use this equation we need to know the probability distribution of inputs. Here, like in Section~\ref{Section:Randomness}, we adopt an assumption of uniformity. We assume that the input expressions are drawn uniformly from a given set $E$ with $|E|=K$ elements. Also, as we know that Boolean expressions indeed describe Boolean functions, we assume that for any Boolean function $F$, which is possible to be an input to the computer, its inverse (i.e., $G=\lnot F$) is also possible to be an input with the same probability. It should be noted that here we consider uniform distributions just as a way of example and, as our argument implies, the approach of using randomness to ameliorate the increase in thermodynamic entropy must work for any other input distribution. 

We first consider the conventional case, where the computer does not use any randomness mechanism. To calculate $H(I|O)$, recall that this is the amount of uncertainty (information) introduced by the input, provided that the output is known. Here the input is a Boolean expression and the output is a satisfying assignment to the Boolean variables. As we have assumed a uniform distribution over $E$, and as we have assumed that for any function $F$, both $F$ and $\lnot F$ are equally likely to be the input, we can conclude that any given variable assignment satisfies half of the Boolean expressions and does not satisfy the other half (Indeed, any assignment that satisfies a Boolean function $F$ obviously does not satisfy the function $G=\lnot F$). Therefore, if the output, i.e., the satisfying variable assignment, is given, the input can only be from that half of the Boolean expressions in the set $E$. Hence we have:
\begin{equation}\label{eq:SAT1} 
H(I|O)= \log_2 (\frac{K}{2}) 
\end{equation}

Regarding $H(O|I)$, as we have assumed that the computer does not exploit randomness and has a macroscopically deterministic behavior, for any input expression the computer gives exactly one specific output without showing any uncertainty in it. This means that $H(O|I)=0$ and in this case we have: 
\begin{equation}\label{eq:SAT_DET} 
\Delta \bar{S}=k_B \ln{2} \cdot (\log_2 (\frac{K}{2})-0)=k_B (\ln{K}-\ln {2}) 
\end{equation}

Now let us move on to the case where randomness is exploited. This time, instead of directly using Eq.~\ref{eq:TM}, we use the chain rule for conditional entropy \cite{Cover}, as indicated in a footnote in Section~\ref{Section:Randomness}, to rewrite Eq.~\ref{eq:TM} as follows:
\begin{equation}\label{eq:TMSimpler} 
\Delta \bar{S}=k_B \ln{2} \cdot [H(I)-H(O)]
\end{equation} 

In Eq.\ \ref{eq:TMSimpler}, $H(I)$ can be viewed as the amount of uncertainty introduced by the input, and since we have assumed a uniform distribution over $E$, we have $H(I)= \log_2 K$. To calculate $H(O)$, assuming that there are $n$ variables, we should note: \emph{i}) There are $2^{n}$ different possible variable assignments, as each variable can be either \textquoteleft False\textquoteright \ or \textquoteleft True\textquoteright, \emph{ii}) Assuming that an expression with $i$ satisfying assignments is given to the computer, by exploiting randomness, any of these $i$ assignments is equally probable to be the output. It should be noted that if the computer did not exploit randomness and behaved deterministically, it would always give one specific assignment as the output, even though there were $i-1$ other acceptable solutions, and \emph{iii}) Due to the uniformity assumptions for the input, the input does not impose any preference for some specific assignments to be the output. Therefore, all variable assignments can be evenly provided at the output. These three items indicate that the output can evenly be any of $2^{n}$ possible assignments and hence we have $H(O)= \log_2 (2^{n})=n$. Therefore, for the case where randomness is exploited, we have:
\begin{equation}\label{eq:SAT_RAN} 
\Delta \bar{S}=k_B \ln{2} \cdot (\log_2 K-n)= k_B (\ln{K}-n\ln {2}) 
\end{equation}

Comparing Eqs.~\ref{eq:SAT_DET} and~\ref{eq:SAT_RAN}, it can be seen that exploiting randomness ameliorates the increase in thermodynamic entropy. The key point here is that this amelioration has been achieved without appending extra information to the output. Indeed, when solving the satisfiability problem, both the computers, i.e., the deterministic (non-random) computer and the computer that exploits randomness, give a string of length $n$ of Ts and Fs as the output (specifying how one should assign values to the $n$ variables of the input expression to satisfy it). The difference is that if one repeatedly uses the deterministic (non-random) computer for the same input, one will always get the same output, but if one repeatedly uses the computer which exploits randomness for the same input, then every time, one can get a different acceptable output, so that all acceptable outputs have equal probability to occur.

At the end of this section, it is worth noting two points: 

\emph{I}) One important point about the use of randomness (macroscopic non-determinism) to ameliorate the increase in thermodynamic entropy is that it can be viewed as an orthogonal technique to reversibility, in the sense that the use of randomness does not limit the influence of reversibility and conversely the use of reversibility does not limit the effectiveness of randomness. This is because these two approaches affect the thermodynamic entropy in different ways, so that, considering Eq.~\ref{eq:TM}, i.e., $\Delta \bar{S}=k_B \ln{2} \cdot [H(I|O)-H(O|I)]$, while reversibility is used to reduce the term $H(I|O)$, randomness is used to increase the term $H(O|I)$. Since the aim of this section was to study randomness, for the sake of clarity and to avoid confusion, we did not exploit reversibility and just focused on randomness, i.e., on increasing $H(O|I)$.

\emph{II}) It was seen that the outputs of the deterministic (non-random) computer and the computer that exploits randomness have the same size (strings of Fs and Ts of a length equal to the number of Boolean variables). Therefore, we can see that the approach of exploiting randomness does not incur extra and redundant bits at the output. Nevertheless, it must be noted that the output of the computer that exploits randomness contains more information than the output of the deterministic computer. Indeed, we know from information theory that it is not just the string length that determines the amount of information; rather it is the level of uncertainty that determines it. For instance, we might have a string of 10 bits but just containing, for example, 2.5 bits of real information due to its relatively low level of uncertainty. We know from information theory that such a 10-bit string can be compressed to a much shorter string with a length close to 2.5 bits (e.g., 3 bits), but in many real-world computing applications such compressions are not performed as they are usually not cost-effective. The proposed approach here is that, for a given output size, if such a gap exists between the actual and the maximum possible amount of information that the output contains, we can include some random information in the output without increasing the output size, thereby increasing $H(O|I)$ in Eq.~\ref{eq:TM}, which results in reducing the thermodynamic entropy. Such a technique not only can be applied for the satisfiability problem and other NP-complete problems, but also can be used for any problem whose output solution contains an amount of information less than the maximum determined by the size of the output.

\section{Daemon computers} \label{Section:Daemon}
So far, we have discussed that randomness (in the sense of macroscopic non-determinism) can be used besides reversibility to ameliorate the increase in thermodynamic entropy in a computation. Now, in this section, we discuss that the use of randomness can apparently lead to a theoretical dilemma, which is of great importance and must be scrutinized. The dilemma is about the second law of thermodynamics. This law, which indicates that for any closed physical system the thermodynamic entropy does not decrease with time \cite{Rex}, is one of the most important and fundamental laws of nature so that any theory about thermodynamic entropy and energy consumption of computing must conform to it. Here we formally state the second law of thermodynamics as:

\begin{axsecond} [Second law of thermodynamics] \label{ax:second}
Let $A$ be a closed physical system and $t_1$ and $t_2$ be two arbitrary time instants so that $t_2>t_1$. Then we have:
\begin{equation}\label{eq:second} 
\Delta \bar{S}_A=\bar{S}_A(t_2)-\bar{S}_A(t_1) \geq \ 0 
\end{equation}
\noindent where $\bar{S}_A(t)$ is the thermodynamic entropy of the system at time $t$. 
\end{axsecond}

It is noteworthy that while our discussion so far has involved thermodynamic entropy, this is the first time in this paper that we aim at considering the second law of thermodynamics, even though thermodynamic entropy and the second law are inherently related\footnote{ Indeed, the concept of thermodynamic entropy was developed by Clausius following the advent of the second law of thermodynamics as a way of modeling the law with mathematical equations \cite{Rex}.}. 

Now let us consider the second law of thermodynamics in the context of computing. We saw in Section~\ref{Section:Randomness} that the change in thermodynamic entropy due to a computation is given by Eq.\ \ref{eq:TM}, i.e.,\ $ \Delta \bar{S}=k_B \ln{2}~\cdot~[H(I|O)-H(O|I)]$, which indeed gives the least entropy change essential for the computation, as we excluded the entropy changes that are unnecessary to the computation (See Section~\ref{Section:Randomness}). When we use a computer with completely deterministic behavior that does not exploit any randomness mechanism, we clearly have $H(O|I)=0$, which means that $H(O|I)$ can be removed from Eq.\ \ref{eq:TM}, turning the equation into $\Delta \bar{S}=k_B \ln{2} \cdot H(I|O)$. In this case, the equation is consistent with the second law of thermodynamics, since we know from information theory \cite{Cover} that $H(I|O)$ cannot be negative and hence we have $\Delta \bar{S} \geq \ 0$. The dilemma can show up when the computer exploits a randomness mechanism as proposed in Sections~\ref{Section:Randomness} and~\ref{Section:NP}. In this case, if $H(O|I)$, i.e., the uncertainty (information) caused by the use of randomness during the computation could be greater than $H(I|O)$, i.e., the lost information during the computation, then the term $H(I|O)- H(O|I)$ in Eq. ~\ref{eq:TM} would be negative, which means that the second law of thermodynamics would be violated. An important case of this is when a computer exploits reversibility, which means that $H(I|O)=0$, turning Eq.\ \ref{eq:TM} into $\Delta \bar{S}=-k_B \ln{2} \cdot H(O|I)$. In this case, any use of randomness to increase $H(O|I)$ results in a negative $\Delta \bar{S}$, thereby violating the second law. One specific instance of this is when we use reversible computers to solve NP-complete problems. Consider a reversible computer that solves the satisfiability problem and for each input, the output not only contains a satisfying assignment but also contains an intact copy of the input expression for the sake of reversibility. For this computer, due to reversibility, we have $H(I|O)=0$. But to obtain $H(O|I)$, as mentioned in Section~\ref{Section:NP}, we need to know the probability distributions of the input and output. In this regard, similar to the assumptions in Section~\ref{Section:NP}, we assume uniform distributions, i.e., we assume that all Boolean functions are equally likely to be given as input\footnote{It is noteworthy that a Boolean function can be represented by different expressions. For example, expressions $p\lor \lnot q$, $\lnot(\lnot p \land q)$, and $q \rightarrow p$ are equivalent and they all represent the same function. Here we have assumed a uniform probability distribution over Boolean functions given by input expressions, without having any assumption for the probability distribution over expressions that are used to represent functions and can be given as input.} and also we assume that all the assignments that satisfy an input are equally likely to be provided at the output. Based on these assumptions, and as shown in Appendix 3, $H(O|I)$ can be formulated as:
\begin{equation}\label{eq:HOI_NP} 
H(O|I)= \frac{1}{2^{2^{n}}} \sum_{k=2}^{2^{n}} {2^{n}\choose k}\cdot \log_2 k
\end{equation}

\noindent where $n$ is the number of variables of the input expressions. It can be easily seen from Eq.~\ref{eq:HOI_NP} that $H(O|I)$ is always positive (note that $n\geq 1$ as $n$ is the number of variables and we assume there is at least one variable in the input expression). Since for such a reversible computer we have $H(I|O)=0$ and $H(O|I) > 0$, the term $H(I|O)- H(O|I)$ in Eq.~\ref{eq:TM} is negative and consequently so is the entropy change $\Delta \bar{S}$, which is a violation of the second law. 

While here we have considered computations that may violate the second law, there have also been some notable thought experiments in the context of thermodynamics and statistical mechanics where the second law can be violated \cite {Bennett3, Leff}. However there is an important difference between these thought experiments and the computational case we discussed here. In the thought experiments, there is a being with an undefined nature, called Maxwell\textquoteright s daemon, whose intervention in a physical system at the molecule level causes the violation. Indeed, since the inception of Maxwell\textquoteright s daemon, its nature has been a topic of intense debate. But in the computational case that we discussed here, a computer which is equipped with a randomness mechanism may violate the second law without any need for the intervention of an undefined being and hence here we name such a computer itself a daemon computer. 

By the use of the theory developed in this paper, we will argue in the rest of this section that randomness (macroscopic non-determinism) not only plays a central role in daemon computers but is also an inseparable part of conventional thought experiments that include Maxwell\textquoteright s daemon, even though the important role of randomness\footnote{It should be emphasized that here, by the term randomness we only mean macroscopic non-determinism and we do not refer to any other issue, such as thermal fluctuations that are commonly considered in thermodynamics or statistical mechanics.} has been overlooked in these thought experiments. Indeed, we state this in the form of the following theorem, which has been proved in Appendix 2. 

\begin{thmSecondLaw} [Randomness and the second law of thermodynamics] \label{thm:SecondLaw}
Suppose that $A$ is a closed physical system where the second law of thermodynamics is violated. Then system $A$ must show a macroscopically non-deterministic (random) behavior.
\end{thmSecondLaw}

The important point is that Theorem~\ref{thm:SecondLaw} is very general and is valid for any closed physical system. Therefore, it must also hold for the closed systems in thought experiments with Maxwell\textquoteright s daemon, which means that randomness, in the sense of macroscopic non-determinism, must exist in their behaviors. For instance, consider Szilard\textquoteright s engine \cite{Szilard}, which is one of the most well-known examples of these thought experiments. In Szilard\textquoteright s engine, a single molecule is confined within a box. In the first step of the engine cycle, a partition is inserted in the middle of the box so that the molecule is trapped in one half of the box. In the next step, the daemon, which controls the operation of the engine, observes the box to detect in which half of the box the molecule is trapped. Here, we do not intend to explain the operation of Szilard\textquoteright s engine and for more information please refer to related publications such as \cite{ Bennett3,Szilard}. But the macroscopic non-determinism (randomness) in the operation of Szilard\textquoteright s engine is apparent from the first two steps that we mentioned. First, the issue of in which half of the box the molecule is trapped must be considered as a macroscopic property of the engine as the daemon can observe and perceive it (see the definitions of perfect macrostate in Section~\ref{Section:Information}). Second, in Szilard\textquoteright s engine, when the partition is inserted, the molecule might be trapped in any half of the box with equal probabilities, and hence the state (the half where the molecule is trapped) is set randomly and non-deterministically. 

In short, we can conclude from the discussion here, especially from Theorem~\ref{thm:SecondLaw}, that randomness in the sense of macroscopic non-determinism is the key point in daemon computers (and Maxwell\textquoteright s daemons) to challenge the second law of thermodynamics. On the other hand, we must note that the second law of thermodynamics is one of the most fundamental laws of nature, whose violation is unacceptable\footnote{ Indeed, any violation of the second law has unbelievable consequences. For example, the second law is the only law in physics which is not time-symmetric in the sense that it differentiates between past and future \cite{Rex}. In other words, without the second law, moving backward in time would be as normal as moving forward. The other unbelievable consequence of violating the second law is that it would allow devising perpetual motion machines \cite{Rex,Leff}, which are imaginary engines that can produce work without actually consuming any fuel. These unbelievable consequences are the reasons make us consider the second law as an unbreakable law of nature.}. Therefore here we come to an interesting riddle which is \textquotedblleft How is it possible that our theory, which already includes randomness (macroscopic non-determinism), can also include (conform with) the second law without incurring any contradiction?\textquotedblright . It is clear that doing this requires rejecting the existence of daemon computers. In the next section, we exploit the theory developed in this paper to provide a possible explanation, which perfectly conforms to the second law. However, instead of daemon computers, this explanation necessitates accepting the existence of another group of strange computers that we give them the name \textquoteleft clairvoyant computers\textquoteright .

\section{Clairvoyant Computers} \label{Section:Clairvoyant} 
As mentioned in Section~\ref{Section:Daemon}, violating the second law of thermodynamics is unacceptable. But then the question is \textquotedblleft How can we address the inconsistency between having daemon computers as an outgrowth of randomness (macroscopic non-determinism) on one hand and the second law of thermodynamics on the other?\textquotedblright

It is noteworthy that traditional Maxwell\textquoteright s daemons, like Szilard\textquoteright s engine, are thought experiments with imaginary apparatuses so that the attempts to achieve consistency with the second law usually involve finding some subtle points in the operation of the respective apparatus (see for example \cite{Bennett3}). However, daemon computers, as introduced in Section~\ref{Section:Daemon}, are completely different from Maxwell's daemons in that daemon computers do not involve any thought experiment or imaginary apparatus with potential tricks or subtle points. Rather we have mathematically shown in Section~\ref{Section:Daemon} that in computers that exploit real randomness (macroscopic non-determinism), the second law can be violated. Here, the riddle is clear-cut without any room for tricks, ambiguity, and vagueness. Likewise, we adopt a mathematical approach to provide a solution to the riddle by achieving consistency with the second law. This is what we will do in the rest of this section by introducing one important theorem that provides an alternative to accepting the existence of daemon computers. 

The following theorem indicates that if a computer with macroscopically non-deterministic behavior (i.e., equipped with a mechanism for real randomness) must conform to the second law of thermodynamics, it must gain information from other physical systems through its randomness mechanism! We name this weird property of \emph{gaining information from other physical systems through real randomness} the property of clairvoyance. The theorem here indeed indicates that if the second law is an unbreakable law of nature (which means daemon computers cannot exist), then we must accept the existence of clairvoyant computers. We have proved this theorem in Appendix 2, but we discuss the concept of the theorem in this section to provide an insight into the issue. It is also noteworthy that this theorem involves the mutual information between two physical systems. Indeed the mutual information between two variables $X$ and $Y$, denoted by $I(X;Y)$, indicates the amount of information that one of these two variables contains about the other one, or in other words, corresponds to the intersection of the information in $X$ with the information in $Y$ \cite{Cover}.

\begin{thmClairvoyance} [The property of clairvoyance] \label{thm:clairvoyance}
Suppose that $A$ is a closed physical system which consists of two smaller systems $B$ and $C$. Assuming that the following three conditions hold for any time interval $t_1$ to $t_2$ ($t_2 > t_1$): 

\begin{itemize}
\itemsep0em
\item[] 1) The system should conform to the second law of thermodynamics, hence we have $\bar{S}_A(t_2) \geq \bar{S}_A(t_1)$.
\item[] 2) System B does not lose any information at the macroscopic level (i.e., it possesses reversibility), hence we have $H(M_B(t_1)|M_B(t_2))=0$. 
\item[] 3) System C has no change in its information content, i.e., does not lose or gain any information. Therefore, we have $H(M_C(t_1)|M_C(t_2))=0$ and $H(M_C(t_2)|M_C(t_1))=0$. 
\end{itemize}

\noindent then during the time interval $t_1$ to $t_2$, we have: 

Part I: If system $B$ has macroscopically non-deterministic behavior (i.e., behaves randomly), which means $H(M_B(t_2)|M_B(t_1))\neq 0$, then it must gain information from system $C$, i.e., we have: 
\begin{equation}\label{eq:Conclusion} 
I(M_B(t_2);M_C(t_2)) > I(M_B(t_1);M_C(t_1))
\end{equation}

Part II: But if system $B$ behaves completely in a deterministic manner (i.e., does not behave randomly), which means $H(M_B(t_2)|M_B(t_1))=0$, then it gains no information from system $C$, i.e., we have: 
\begin{equation}\label{eq:Conclusion2} 
I(M_B(t_2);M_C(t_2)) = I(M_B(t_1);M_C(t_1))
\end{equation}
\end{thmClairvoyance}

The proof for Theorem~\ref {thm:clairvoyance} uses Lemma~\ref{lm:removeREDApp}, but for the sake of brevity this lemma has not been provided in the body of the paper and can be found in Appendix 1 including its proof. There are important points about this theorem that should be noted: 

\begin{itemize}
\item When system $B$ behaves randomly, the mutual information between systems $B$ and $C$ increases because system $B$ gains information about system $C$, and not because system $C$ gains information about system $B$. Indeed, based on condition 3, the information content of system $C$ remains unchanged during the respective time interval and hence it gains no information. 

\item The information that system $B$ gains about system $C$ must be received through the use of randomness in system $B$. This is because the information gain has happened in Part I, where the randomness is exploited, and in Part II, where there is no randomness, there is no information gain. 

\item It must be emphasized that, as it can be seen from the statement of Theorem~\ref{thm:clairvoyance} and its proof in Appendix 2, the property of clairvoyance is a direct consequence of the second law of thermodynamics. If the second law did not have to be applied, we would not be obliged to accept the property of clairvoyance and, instead, we could accept the existence of daemon computers.
\end{itemize}

One interesting and prominent issue about the theorem of clairvoyance (Theorem~\ref{thm:clairvoyance}) is that it presumably provides a theoretical explanation for the observation made in The Global Consciousness Project \cite{GCPP} in Princeton University. In this project, it has been observed that values from real random number generators show correlation with major events. The current explanation for this observation (see \cite{GCPP}) is that major events synchronize the feelings of millions of people and people\textquoteright s coherent consciousness affects the behavior of random number generators via some subtle links. But here in this paper, another possibility has been propounded, which is the possibility that the reason behind this strange observation in the behavior of random number generators is the second law of thermodynamics as indicated by Theorem~\ref{thm:clairvoyance}.

\section{Conforming to existing theories} \label{Section:Conforming} 

In this section, we discuss the important issue of how the theory in this paper conforms to widely accepted existing theories. In this regard, the first thing that we must note is that all the three axioms in this paper are indeed well-known laws that have been adopted from existing well-established theories. Let us have a quick overview: 

\noindent \emph{Axiom~\ref{ax:inf}} is the law of information conservation and is valid due to Liouville's theorem \cite[Lecture 1]{Susskind}. 

\noindent \emph{Axiom~\ref{ax:mic}} states that each microstate corresponds with exactly one macrostate (even though the inverse is not true as each macrostate corresponds to many microstates), which is a well-known premise in statistical mechanics \cite{Pathria,Rex}. 

\noindent \emph{Axiom~\ref{ax:second}} is the second law of thermodynamics and probably the most well-known among the three axioms \cite{Feynman,Pathria,Rex}. 

The only thing that might be considered as slightly new is Postulate~\ref{pst:main}, stating that the thermodynamic entropy of a physical system $A$ is always given by $H(\mathfrak{M}_A(t)|M_A(t))$ without requiring any precondition like thermal equilibrium. But as mentioned previously, there are at least two strong evidences for the validity of this postulate. One of them is Theorem~\ref{thm:ent}, which indeed indicates that Postulate~\ref{pst:main} becomes a provable theorem within the axiomatic framework of classical thermodynamics and statistical mechanics. Therefore, this postulate can be considered as a generalization of what is already accepted in well-established theories to beyond the framework of those theories (more accurately when thermal equilibrium is not assumed). The other evidence is the property of additivity which will be discussed in this section. Additivity is indeed one of the most important properties of thermodynamic entropy within the framework of classical thermodynamics and statistical mechanics and we will provide a theorem in this section indicating that the generalization provided by Postulate~\ref{pst:main} still preserves the property of additivity, even outside that framework, i.e., where assumptions such as thermal equilibrium are not required.

We begin the discussion by providing a theorem about the information that different physical systems bear about each other. The theorem is indeed a mathematical formulation of the fact that the only discernible information from the perfect microstate of a physical system is what can be discerned from its perfect macrostate. We name this theorem the property of indiscernibility.
 
\begin{thmindiscernibility} [The property of indiscernibility] \label{thm:Indiscernibility}
 
Suppose that A and B are two separate physical systems, i.e., neither of them is a part or component of the other one, then we have: 
\begin{equation}\label{eq: Indiscernibility} 
H(\mathfrak{M}_A(t)|\mathfrak{M}_B(t))= H(\mathfrak{M}_A(t)| M_B(t))
\end{equation}

\end{thmindiscernibility}

This theorem has been proved in Appendix 2. 

Now, only based on this theorem, i.e., the property of indiscernibility, and without any other assumption, we prove an important theorem about the mutual information between two physical systems. 

\begin{thmMutual} [Mutual information between two physical systems] \label{thm:Mutual}  

Let $A$ and $B$ be two physical systems. Then based on the property of indiscernibility as the only assumption we have:
\begin{equation}\label{eq:Mutual} 
I(\mathfrak{M}_A(t);\mathfrak{M}_B(t))=I(\mathfrak{M}_A(t);M _B(t))=I(M_A(t);\mathfrak{M}_B(t))=I(M_A(t);M_B(t))
\end{equation}
\end{thmMutual} 

The proof for Theorem~\ref{thm:Mutual} is rather complicated and uses Lemmata~\ref{lm:removeREDApp} and~\ref{lm:MoreCond}. For the sake of brevity, these lemmata have not been provided in the body of the paper and can be found in Appendix 1, including their proofs. 

One important theorem, which can be proved from Theorem~\ref{thm:Mutual}, is that the conditional entropy $H(\mathfrak{M}(t)|M(t))$, which is postulated (Postulate~\ref{pst:main} in Section~\ref{Section:Information}) to give the thermodynamic entropy of a system, is additive. Here, we prove that this property holds for $H(\mathfrak{M}(t)|M(t))$ without requiring assumptions such as thermal equilibrium or equal a priori probabilities \cite{Pathria,Rex} and only with assuming the property of indiscernibility\footnote{ Indeed, we prove the property of additivity only based on Theorem~\ref{thm:Mutual}, which in turn is proved only based on the property of indiscernibility, i.e., Theorem~\ref{thm:Indiscernibility}.}. It is also noteworthy that, as it can be seen from the proof of Theorem~\ref{thm:Indiscernibility} in Appenidx 2, the property of indiscernibitliy comes only from the definitions of microstate and macrostate (of course in their perfect sense, i.e., Definitions \ref{def:micro} and \ref{def:macro}) and does not require any other assumption. 

\begin{thmAdditivity} [The property of additivity] \label{thm:additivity}
Let $A$ and $B$ be two physical systems. Then by only assuming the property of indiscernibility and without any other assumption, we have: 
\begin{equation}\label{eq:additivity} 
H(\mathfrak{M} _A(t), \mathfrak{M} _B(t)|M_A(t), M_B(t))= H(\mathfrak{M} _A(t)|M_A(t))+ H(\mathfrak{M} _B(t)|M_B(t))
\end{equation}
\end{thmAdditivity}

It is worth emphasizing that this theorem, which states that $H(\mathfrak{M}(t)|M(t))$ is additive, is another strong evidence for Postulate~\ref{pst:main} (aside from Theorem~\ref{thm:ent} in Section~\ref{Section:Information}). Indeed, our only reason to put forward Postulate~\ref{pst:main} is to propound that, for any physical system, $H(\mathfrak{M}(t)|M(t))$ gives the thermodynamic entropy in such a general sense that does not require any assumption. Otherwise, we would not need to put forward the postulate because we have proved in Theorem~\ref{thm:ent} that $H(\mathfrak{M}(t)|M(t))$ gives the thermodynamic entropy within the axiomatic framework where thermodynamic entropy is traditionally defined (Note that classical thermodynamics and statistical mechanics are based on assumptions such as thermal equilibrium and equal a priori probabilities \cite{Pathria}). Since additivity is one of the most important properties of thermodynamic entropy and since Theorem~\ref{thm:additivity} shows that $H(\mathfrak{M}(t)|M(t))$ possesses this property without having the traditional assumptions of thermodynamics and statistical mechanics, Theorem~\ref{thm:additivity} is a strong evidence for the generalization that Postulate~\ref{pst:main} offers. 

\section{Summary} \label{Section:Summary} 
This paper is an effort towards forming a formal and mathematical theory to explain energy consumption of computing. It has a highly interdisciplinary nature, involving computer science, thermodynamics, statistical mechanics, and information theory. We have proved several important theorems, resulting in some notable conclusions. For instance, one important conclusion is that real randomness might be intentionally exploited to go beyond the energy saving achievable from using reversible computing on its own. It is an important conclusion because while reversible computing has attracted so much attention in previous work on low energy computing, this conclusion indicates that real randomness can even be more useful than reversibility in achieving low energy consumption. This has been especially shown for the case of NP-complete problems. As another instance, the theory implies that in order to conform to the second law of thermodynamics, some computers must exist that can gain information about other physical systems through exploiting randomness. This conclusion is very important, as it presumably provides a theoretical explanation for a notable practical observation made in a project at Princeton University called the global consciousness project.
Just like any other theory, the theory in this paper is not limited to the matters discussed and the conclusions made in this paper. The reasoning in this paper can be extended much further, obtaining other important conclusions and theorems. Nevertheless, obviously this paper does not have enough room to discuss how the theory can be extended.
\\~\\~\\

\noindent {\Large {Appendix 1: Information-Theoretical Lemmata}}

\begin{lmfuncApp} [Information theoretical notion of functions] \label{lm:funcApp}
Variable $X$ uniquely determines the value of variable $Y$ (or in other words $Y$ is a function of $X$) if and only if the conditional entropy $H(Y|X)$ is equal to zero.
\end{lmfuncApp}

\noindent \textbf{Proof:}  This lemma can be written as $H(Y|X)=0 \iff \exists f: Y=f(X)$. Since this is in the if-and-only-if form, the proof has two parts: 
\\~\\
\noindent Part 1: [$\exists f: Y=f(X)] \Longrightarrow H(Y|X)=0$, i.e., if there exists a function $f$ so that $Y=f(X)$ then $H(Y|X)=0$. 

To prove this, we need to consider that conditional entropy is defined as \cite{Cover}: 
\begin{equation}\label{eq:condent} 
H(Y|X)=\underset{x}{\mathbb{E}}[H(Y|X=x)]=\sum_{x\in D_X} [P(X=x)H(Y|X=x)]
\end{equation}
 
\noindent where $D_X$ is the universe of $X$ (i.e., $X$ may take any value from $D_X$ and cannot take any value which is not in $D_X$) and $H(Y|X=x)$ is: 
\begin{equation}\label{eq:condent2} 
H(Y|X=x)=-\sum_{y\in D_Y} [P(Y=y|X=x)\log_2 P(Y=y|X=x)]
\end{equation}

Noting that there exist a function $f$ so that $Y=f(X)$, we consider two possible cases for $P(Y=y|X=x)$, namely either $y=f(x)$ or $y\neq f(x)$. As $Y=f(X)$, in the case $y=f(x)$, we are sure that by having $X=x$, we must have $Y=y$, which means $P(Y=y|X=x)=1$. Also, as $Y=f(X)$, in the case $y\neq f(x)$, we are sure that by having $X=x$ we must have $Y\neq y$, which means $P(Y=y|X=x)=0$. In both cases, $P(Y=y|X=x)\log_2 P(Y=y|X=x)$ is equal to zero\footnote{It should be noted that $p\log p$ is equal to $0$ when $p=0$ by the definition of entropy \cite[page~14]{Cover}.}. Therefore, in Eq.~\ref{eq:condent2}, for all values of $y\in D_Y$, we have $P(Y=y|X=x)\log_2 P(Y=y|X=x)=0$ and hence the summation over it is also $0$, i.e., we have $H(Y|X=x)=0$. Since $H(Y|X=x)=0$, regardless of the value of $x$, we conclude from Eq.~\ref{eq:condent} that $H(Y|X)=0$ and the first part of the lemma is proved. 
\\~\\
\noindent Part 2: $H(Y|X)=0 \Longrightarrow \exists f: Y=f(X)$, i.e., If $H(Y|X)=0$, then there exists a function $f$ so that $Y=f(X)$.

We know that $-p\log p$ is non-negative as $0\le p\le 1$. Therefore, the value of $H(Y|X=x)$ given by Eq.~\ref{eq:condent2} is also non-negative. On the other hand, from the assumption we know that $H(Y|X)=0$ and hence by definition (given by Eq.~\ref{eq:condent}) we have: 
\begin{equation} \label{eq:sumzero} 
\sum_{x\in D_X} [P(X=x)H(Y|X=x)]=0
\end{equation} 

Therefore, noting that $P(X=x)\neq 0$ (as $X$ may take any value from its universe $D_X$), since $H(Y|X=x)$ cannot be negative in Eq.~\ref{eq:sumzero}, we must have $H(Y|X=x)=0$. Hence, we have:
\begin{equation}\label{eq:allH1} 
\forall x\in D_X: H(Y|X=x)=0
\end{equation}

Substituting for $H(Y|X=x)$ in Eq.~\ref{eq:allH1} from Eq.~\ref{eq:condent2} (i.e., the definition for $H(Y|X=x)$), we have:
\begin{equation}\label{eq:allH2} 
\forall x\in D_X: -\sum_{y\in D_Y}P(Y=y|X=x)\log_2 P(Y=y|X=x)=0
\end{equation}

Since $-p\log p$ is non-negative, we conclude from Eq.~\ref{eq:allH2} that:
\begin{equation}\label{eq:2all} 
\forall x\in D_X, \forall y\in D_Y: P(Y=y|X=x)\log_2 P(Y=y|X=x)=0
\end{equation}

\noindent which means:
\begin{equation}\label{eq:2all2} 
\forall x\in D_X, \forall y\in D_Y: P(Y=y|X=x)=1 \quad \textrm{or} \quad P(Y=y|X=x)=0
\end{equation}

However, considering that the sum of the probabilities of all possibilities must be 1, namely in this case, $\sum_{y\in D_Y}P(Y=y|X=x)=1$, we conclude from Eq.~\ref{eq:2all2} that for each value of $x\in D_X$, there is only one value of $y\in D_Y$ for which we have $P(Y=y|X=x)=1$ and for all other values of $y$ we have $P(Y=y|X=x)=0$. That is to say, for each value $x$ of $X$ ($x\in D_X$), there is exactly one possible value for $Y$, denoted by $y$ ($y \in D_Y$). By noting the definition of functions in mathematics, this means that $Y$ is a function of $X$.

As we have proved both parts, the lemma is proved. $\blacksquare$

\begin{lmremoveApp} [Removing non-contributing variables] \label{lm:removeApp} 
Assuming that variables $W$ and $Y$ bear no information \emph{(}i.e., $H(W)=0$ and $H(Y)=0$\emph{)}, we have $H(W,X|Y,Z)=H(X|Z)$. 
\end{lmremoveApp}

\noindent \textbf{Proof:} We first prove that for any two variables $A$ and $B$, we have:
\begin{equation}\label{eq:SimpleImplication} 
H(A)=0 \quad \Rightarrow \quad H(A|B)=0
\end{equation}

To show this, it should be noted that it has been proved in information theory that conditioning reduces entropy \cite{Cover}, i.e., $H(A|B) \leqslant H(A)$. On the other hand, we have the assumption $H(A)=0$ and hence $H(A|B) \leqslant 0$. However, we know from information theory that $H(A|B)$ cannot be negative \cite{Cover}, therefore we must have $H(A|B)=0$ and Eq.~\ref{eq:SimpleImplication} is proved. Now using Eq.~\ref{eq:SimpleImplication}, we can simply prove:
\begin{equation}\label{eq:SimpleImplication2} 
H(A)=0 \quad \Rightarrow \quad H(A,B)=H(B)
\end{equation}

For this purpose, we use the chain rule in information theory \cite{Cover}, which states that $H(A,B)=H(A|B)+H(B)$. But based on Eq.~\ref{eq:SimpleImplication} we know $H(A|B)=0$, therefore we have $H(A,B)=H(B)$ and Eq.~\ref{eq:SimpleImplication2} is proved too. 

Now, we can use Eq.~\ref{eq:SimpleImplication2} to prove the lemma, i.e., we show, assuming $H(W)=0$ and $H(Y)=0$, that we have $H(W,X|Y,Z)=H(X|Z)$. To do this, we first use the chain rule to write: 
\begin{equation}\label{eq:Step1} 
H(W,X|Y,Z)=H(W,X,Y,Z)-H(Y,Z)
\end{equation}

However, as $H(W)=0$ and $H(Y)=0$, we conclude from Eq.~\ref{eq:SimpleImplication2} that $H(W,X,Y,Z)=H(X,Z)$ and $H(Y,Z)=H(Z)$. If we substitute these into Eq.~\ref{eq:Step1} and use the chain rule, we obtain: 
\begin{equation}\label{eq:Step2} 
H(W,X|Y,Z)=H(X,Z)-H(Z)=H(X|Z)
\end{equation}

\noindent and the lemma is proved. $\blacksquare$

\begin{lmremoveREDApp} [Removing redundant variables] \label{lm:removeREDApp}
If the variable $Y$ is a function of the variable $X$, then the variable $Y$ in the expressions $H(X,Y)$, $H(X,Y|Z)$, $H(Z|X,Y)$, and $I(X,Y;Z)$ is redundant and can be removed, i.e., we have: 
\begin{equation} \label{eq:RM}
H(Y,X)=H(X), \quad H(X,Y|Z)=H(X|Z), \quad H(Z|X,Y)=H(Z|X), \quad I(X,Y;Z)=I(X;Z)
\end{equation}
\end{lmremoveREDApp} 

\noindent \textbf{Proof:} We know from the chain rule in information theory \cite{Cover} that $H(Y,X)=H(Y|X)+H(X)$. On the other hand, the assumption of the lemma states that $Y$ is a function of $X$ and in this case, based on Lemma~\ref{lm:funcApp} we have $H(Y|X)=0$, hence if we substitute this into the equation of the chain rule, we obtain: 
\begin{equation} \label{eq:RM1}
H(Y,X)=H(X) 
\end{equation}

\noindent and the first equation of the lemma is proved. 

In information theory, the chain rule for conditional entropy states that \cite{Cover}: 
\begin{equation} \label{eq:Intermed1}
H(Y,X|Z)=H(Y|X,Z)+H(X|Z) 
\end{equation}

However, since $Y$ is a function of $X$, based on Lemma~\ref{lm:funcApp} we have $H(Y|X,Z)=0$ and hence if we substitute this into Eq.~\ref{eq:Intermed1}, we obtain: 
\begin{equation} \label{eq:RM2}
H(Y,X|Z)=H(X|Z) 
\end{equation}

\noindent and the second equation of the lemma is proved. By adding $H(Z)$ to both sides of Eq.~\ref{eq:RM2} and based on the chain rule, we obtain: 
\begin{equation} \label{eq:Intermed2}
H(Y,X,Z)=H(X,Z) 
\end{equation}

From Eqs.~\ref{eq:RM1} and~\ref{eq:Intermed2}, we conclude: 
\begin{equation} \label{eq:Intermed3}
H(Y,X,Z)-H(Y,X)=H(X,Z)-H(X) 
\end{equation}

\noindent that, based on the chain rule, can be rewritten as: 
\begin{equation} \label{eq:RM3}
H(Z|X,Y)=H(Z|X) 
\end{equation}
\noindent and the third equation of the lemma is proved. To prove the last equation, we know from information theory that the mutual information $I(X,Y;Z)$ can be written as \cite{Cover}:
\begin{equation} \label{eq:Intermed4}
I(X,Y;Z)=H(X,Y)-H(X,Y|Z) 
\end{equation}

But if we substitute from Eqs.~\ref{eq:RM1} and~\ref{eq:RM2} into Eq. ~\ref{eq:Intermed4}, we have: 
\begin{equation} \label{eq:Intermed5}
I(X,Y;Z)=H(X)-H(X|Z) 
\end{equation}

We know from information theory that the right hand side of this equation is $I(X;Z)$ \cite{Cover}. Therefore we have: 
\begin{equation} \label{eq:RM4}
I(X,Y;Z)=I(X;Z) 
\end{equation}

\noindent and the last equation of the lemma is also proved. $\blacksquare$

\begin{lmMoreCondApp} [The more we know, the less uncertainty] \label{lm:MoreCond}
For any three variables $X$, $Y$, and $Z$, we have: 
\begin{equation} \label{eq:MoreCond}
H(Z|X,Y) \leq H(Z|X)
\end{equation}
\end{lmMoreCondApp} 

\noindent \textbf{Proof:} In information theory, the chain rule for mutual information states that \cite{Cover}: 
\begin{equation} \label{eq:Chain4MI}
I(X,Y;Z)=I(X;Z)+I(Y;Z|X)
\end{equation}

On the other hand, we know that conditional mutual information cannot be negative \cite{Cover}, i.e., $I(Y;Z|X) \geq 0$, therefore from Eq.~\ref{eq:Chain4MI} we conclude that:
\begin{equation} \label{eq:IneqInfo}
I(X,Y;Z) \geq I(X;Z)
\end{equation}

From information theory, we also know that for any two variables $A$ and $B$ we have $I(A;B)=H(A)+H(B)-H(A,B)$ \cite{Cover}. Hence, Inequality~\ref{eq:IneqInfo} can be rewritten as:
\begin{equation} \label{eq:IneqInfo2}
H(X,Y)+H(Z)-H(X,Y,Z) \geq H(X)+H(Z)-H(X,Z)
\end{equation}

By subtracting $H(Z)$ from both sides of Inequality~\ref{eq:IneqInfo2} and also multiplying them by $-1$, we obtain: 
\begin{equation} \label{eq:IneqInfo3}
H(X,Y,Z)-H(X,Y) \leq H(X,Z)-H(X)
\end{equation}

By the chain rule \cite{Cover}, we have $H(A|B)=H(A,B)-H(B)$ and hence Inequality~\ref{eq:IneqInfo3} can be written as: 
\begin{equation} \label{eq:IneqFin}
H(Z|X,Y) \leq H(Z|X)
\end{equation}

\noindent and the lemma is proved.  $\blacksquare$
\\~\\~\\

\pagebreak
\noindent {\Large {Appendix 2: Theorems}} 
\begin{thmconstApp} [Constant amount of substantive information] \label{thm:constApp}
Let $A$ be a closed physical system. Then the amount of information represented by the system perfect microstate remains constant with time, i.e., $H(\mathfrak{M}_A(t))=Constant$.
\end{thmconstApp}

\noindent \textbf{Proof:} To prove this theorem, we use the chain rule, which states that for any pair of variables $X$ and $Y$ we have $H(X,Y)=H(X)+H(Y|X)$ \cite{Cover}. Using this rule, for any two time instants $t_1$ and $t_2$, we can write: 
\begin{equation}\label{eq:chain1} 
H(\mathfrak{M}_A(t_1),\mathfrak{M}_A(t_2))=H(\mathfrak{M}_A(t_1))+H(\mathfrak{M}_A(t_2)|\mathfrak{M}_A(t_2)) 
=H(\mathfrak{M}_A(t_2))+H(\mathfrak{M}_A(t_1)|\mathfrak{M}_A(t_1))
\end{equation}

However, from the law of information conservation (i.e., Eq.~\ref{eq:infconvfull}) we know $H(\mathfrak{M}_A(t_2)|\mathfrak{M}_A(t_1))=H(\mathfrak{M}_A(t_1)|\mathfrak{M}_A(t_2))=0$, hence, Eq.~\ref{eq:chain1} can be written as: 
\begin{equation}\label{eq:removedApp} 
H(\mathfrak{M}_A(t_1),\mathfrak{M}_A(t_2))=H(\mathfrak{M}_A(t_1))=H(\mathfrak{M}_A(t_2))
\end{equation}

Therefore, in short, we have: 
\begin{equation}\label{eq:removedApp2} 
\forall t_1,t_2: H(\mathfrak{M}_A(t_1))=H(\mathfrak{M}_A(t_2))
\end{equation}

\noindent and the theorem is proved. $\blacksquare$

\begin{thmentropyApp} [Information theoretical notion of thermodynamic entropy] \label{thm:entApp}
Let $A$ be a physical system. Within the axiomatic framework where thermodynamic entropy is defined, we have:
\begin{equation}\label{eq:thmentApp} 
H(\mathfrak{M}_A(t)|M_A(t))=\frac{1}{k_B \ln{2}}\bar{S}_A(t)
\end{equation}

\noindent where $k_B$ is Boltzmann\textquoteright s constant and $\bar{S}_A(t)$ is the thermodynamic entropy of system $A$ at time $t$.
\end{thmentropyApp}

\noindent \textbf{Proof:} 
Within the axiomatic framework where thermodynamic entropy is defined, it has been proved that for each macrostate $m$ of a physical system, say system $A$, the thermodynamic entropy of the system is equal to \cite{Susskind,Pathria}: 
\begin{equation}\label{eq:Gibbs} 
S_A(m)=-k_B \sum_{i \in \Gamma (m)}p_i\cdot \ln{p_i}
\end{equation}

\noindent where $k_B$ is Boltzmann\textquoteright s constant, $\Gamma(m)$ is the set of all microstates corresponding to the macrostate $m$, and $p_i$ is the probability of finding the system in the microstate $i$ given that the system is in the macrostate $m$. It should be noted that in Eq.~\ref{eq:Gibbs}, $p_i$ is indeed a conditional probability as it is known that the system is in the macrostate $m$, i.e., we have: 
\begin{equation}\label{eq:Text} 
p_i=P(\textrm{The system is in the microstate} \ i|\ \textrm{The system is in the macrostate} \ m)
\end{equation}

Throughout this paper, we denote the microstate and macrostate of a physical system $A$ at time $t$ by $\mathfrak{M}_A(t)$ and $M_A(t)$ respectively. Therefore, considering that Eq.~\ref{eq:Text} holds at any time, we can rewrite Eq.~\ref{eq:Text} in a more formal way as:
\begin{equation}\label{eq:CondFormal} 
p_i=P(\mathfrak{M}_A(t)=i|M_A(t)=m)
\end{equation}

If we substitute for $p_i$ from Eq.~\ref{eq:CondFormal} into Eq.~\ref{eq:Gibbs}, we have:
\begin{equation}\label{eq:GibbsFull} 
S_A(m)=-k_B \sum_{i \in \Gamma (m)}{[P(\mathfrak{M}_A(t)=i|M_A(t)=m)\cdot \ln{P(\mathfrak{M}_A(t)=i|M_A(t)=m)}]}
\end{equation}

Now, let us turn to the information theoretic term $H(\mathfrak{M}_A(t)|M_A(t)=m)$. By definition (see Eq.~\ref{eq:condent2}), we have: 
\begin{equation}\label{eq:ThrEnt1} 
H(\mathfrak{M}_A(t)|M_A(t)=m)=-\sum_{i \in \Gamma} [P(\mathfrak{M}_A(t)=i|M_A(t)=m)\log_2 P(\mathfrak{M}_A(t)=i |M_A(t)=m)]
\end{equation}

\noindent where $\Gamma$ is the set of all microstates. However, we must note that in Eq.~\ref{eq:ThrEnt1}, for all $i\notin \Gamma (m)$, if the system macrostate is $M_A(t))=m$, it is impossible that the system microstate can be $\mathfrak{M}_A(t)=i$. This is indeed due to Axiom~\ref{ax:mic} which states that each macrostate $m$ only corresponds to microstates that comprise the set $\Gamma (m)$. Therefore, we have: 
\begin{equation}\label{eq:Part1} 
P(\mathfrak{M}_A(t)=i|M_A(t)=m)=0 \quad \textrm{when}\ i \notin \Gamma (m)
\end{equation}

Hence, in Eq.~\ref{eq:ThrEnt1}, the sum can be taken over $\Gamma (m)$ instead of $\Gamma$, which means Eq.~\ref{eq:ThrEnt1} can be written as:
\begin{equation}\label{eq:ThrEnt2} 
H(\mathfrak{M}_A(t)|M_A(t)=m)=-\sum_{i \in \Gamma (m)} [P(\mathfrak{M}_A(t)=i |M_A(t)=m)\log_2 P(\mathfrak{M}_A(t)=i |M_A(t)=m)]
\end{equation}

Now, if we compare Eqs.~\ref{eq:GibbsFull} and~\ref{eq:ThrEnt2}, we can easily conclude that: 
\begin{equation}\label{eq:ThrEnt4} 
H(\mathfrak{M}_A(t)|M_A(t)=m) = \frac{1}{k_B \ln{2}}S_A(m)
\end{equation}

While $S_A(m)$ is the system thermodynamic entropy when the system macrostate is $m$, the exact macrostate of a physical system might not be known at a given time $t$. In general, the macrostate of a physical system $A$  at a time $t$, denoted by $M_A(t)$, can be any of several possible macrostates with different probabilities. In this case, the thermodynamic entropy of the system at time $t$, denoted by $\bar{S}_A(t)$, is obtained by averaging over possible macrostates at the time $t$, i.e., we have\footnote{A notable example of averaging thermodynamic entropy over possible macrostates is the seminal paper \cite{Szilard}.}:
\begin{equation}\label{eq:ExpEnt} 
\bar{S}_A(t)=\underset{m}{\mathbb{E}}[S_A(M_A(t))]
\end{equation}

Now, by taking expected values of both sides of Eq.~\ref{eq:ThrEnt4} with respect to the macrostate $M_A(t)$  as a random variable, we have:
\begin{equation}\label{eq:ExpSides} 
\underset{m}{\mathbb{E}}[H(\mathfrak{M}_A(t)|M_A(t)=m)] = \frac{1}{k_B \ln{2}}~\underset{m}{\mathbb{E}}[S_A(M_A(t))]
\end{equation}

The left hand side of Eq.~\ref{eq:ExpSides} is indeed $H(\mathfrak{M}_A(t)|M_A(t))$ (see Eq.~\ref{eq:condent}) and its right hand side is given by Eq.~\ref{eq:ExpEnt}. Hence, we reach Eq.~\ref{eq:thmentApp}, i.e.:
\begin{equation} 
H(\mathfrak{M}_A(t)|M_A(t))=\frac{1}{k_B \ln{2}}\bar{S}_A(t)
\end{equation}

\noindent and the theorem is proved. $\blacksquare$

\begin{thmMicroMacroApp} [Relationship between thermodynamic entropy and perceptible information] \label{thm:micmacApp}
Let $A$ be a closed physical system. Then at any time instant $t$, for the perceptible information $H(M_A(t))$ we have: 
\begin{equation}\label{eq:thmmicmacApp} 
C_{SI}=H(M_A(t))+\frac{1}{k_B \ln{2}}\bar{S}_A(t)
\end{equation}

\noindent where $C_{SI}$ is the substantive information of the closed system and is a constant value (see Theorem \ref{thm:const}), $k_B$ is Boltzmann\textquoteright s constant, and $\bar{S}_A(t)$ is the thermodynamic entropy of the system at time $t$.
\end{thmMicroMacroApp}

\noindent \textbf{Proof:} 
To prove this theorem, we use the chain rule in information theory, which states that for any pair of variables $X$ and $Y$, we have $H(X,Y)=H(X)+H(Y|X)$ \cite{Cover}. Using this rule, at any time instant $t$, we have: 
\begin{equation}\label{eq:chain_micmac} 
H(\mathfrak{M}_A(t),M_A(t))=H(\mathfrak{M}_A(t))+H(M_A(t)|\mathfrak{M}_A(t)) 
=H(M_A(t))+H(\mathfrak{M}_A(t)|M_A(t))
\end{equation}

However, we know from Axiom~\ref{ax:mic} (Eq.~\ref{eq:micmac}) that $H(M_A(t)|\mathfrak{M}_A(t))=0$, and therefore we can conclude from Eq.~\ref{eq:chain_micmac} that: 
\begin{equation}\label{eq:Landauer_Original} 
H(\mathfrak{M}_A(t))=H(M_A(t))+H(\mathfrak{M}_A(t)|M_A(t))
\end{equation}

We know from Theorem~\ref{thm:const}  that since $A$ is a closed system, the substantive information $H(\mathfrak{M}_A(t))$ is constant with time; let it be denoted by $C_{SI}$. Also, we know from Postulate~\ref{pst:main} that the conditional entropy $H(\mathfrak{M}_A(t)|M_A(t))$ gives the thermodynamic entropy of the system so that we can use Eq.~\ref{eq:thment} to substitute for it. Therefore, we can write Eq.~\ref{eq:Landauer_Original} as:
\begin{equation}\label{eq:Landauer_Final} 
C_{SI}=H(M_A(t))+\frac{1}{k_B \ln{2}}\bar{S}_A(t)
\end{equation}

\noindent and the theorem is proved. $\blacksquare$
\begin{thmLandauerApp} [The relationship between thermodynamic entropy, reversibility and determinism] \label{thm:landauerApp} 
Let $A$ be a closed physical system, and let $t_1$ and $t_2$ be two arbitrary time instants. Then we have: 
\begin{equation}\label{eq:landauerApp} 
\Delta \bar{S}=\bar{S}(t_2)-\bar{S}(t_1)=k_B \ln{2} \cdot [H(M_A(t_1)|M_A(t_2))-H(M_A(t_2)|M_A(t_1))]
\end{equation}
\noindent where $k_B$ is Boltzmann\textquoteright s constant, and $\bar{S}_A(t)$ is the thermodynamic entropy of the system at time $t$.
\end{thmLandauerApp}

\noindent \textbf{Proof:} 
To prove this theorem, we use the chain rule in information theory, which states that for any pair of variables $X$ and $Y$, we have $H(X,Y)=H(X)+H(Y|X)$ \cite{Cover}. Using this rule, for any two time instants $t_1$ and $t_2$, we can write: 
\begin{equation}\label{eq:chain_macmac} 
H(M_A(t_1),M_A(t_2))=H(M_A(t_1))+H(M_A(t_2)|M_A(t_1))=H(M_A(t_2))+H(M_A(t_1)|M_A(t_2))
\end{equation}

\noindent and it can be easily concluded from Eq.~\ref{eq:chain_macmac} that:
\begin{equation}\label{eq:rephrased_macmac} 
H(M_A(t_1))-H(M_A(t_2))=H(M_A(t_1)|M_A(t_2))-H(M_A(t_2)|M_A(t_1))
\end{equation}

On the other hand, from Theorem~\ref{thm:micmacApp} (Eq.~\ref{eq:thmmicmacApp}), we know that for any two time instants $t_1$ and $t_2$ we have: 
\begin{equation}\label{eq:thmmicmacTWO} 
C_{SI}=H(M_A(t_1))+\frac{1}{k_B \ln{2}}\bar{S}_A(t_1)=H(M_A(t_2))+\frac{1}{k_B \ln{2}}\bar{S}_A(t_2)
\end{equation}

\noindent and it can be easily concluded from Eq.~\ref{eq:thmmicmacTWO} that:
\begin{equation}\label{eq:rephrased_micmacTWO} 
H(M_A(t_1))-H(M_A(t_2))=\frac{1}{k_B \ln{2}} (\bar{S}_A(t_2)-\bar{S}_A(t_1))
\end{equation}

Now, if we compare Eqs.~\ref{eq:rephrased_macmac} and~\ref{eq:rephrased_micmacTWO}, we can easily conclude that:
\begin{equation}\label{eq:landauerAppFinal} 
\bar{S}(t_2)-\bar{S}(t_1)=k_B \ln{2} \cdot [H(M_A(t_1)|M_A(t_2))-H(M_A(t_2)|M_A(t_1))]
\end{equation}

\noindent and the theorem is proved. $\blacksquare$

\begin{thmSecondLawApp} [Randomness and the second law of thermodynamics] \label{thm:SecondLawApp}
Suppose that $A$ is a closed physical system where the second law of thermodynamics is violated. Then system $A$ must show a macroscopically non-deterministic (random) behavior.
\end{thmSecondLawApp}

\noindent \textbf{Proof:} Consider Eq.~\ref{eq:landauer} of Theorem~\ref{thm:landauer}, i.e., 

\begin{equation*}
\Delta \bar{S}=\bar{S}(t_2)-\bar{S}(t_1)=k_B \ln{2} \cdot [H(M_A(t_1)|M_A(t_2))-H(M_A(t_2)|M_A(t_1))]
\end{equation*}

\noindent , which is a general equation valid for all closed physical systems. From this equation, we can easily conclude that as time goes from $t_1$ to $ t_2$, for contradicting the second law, i.e., $\Delta \bar{S} <0$, we must have: 
\begin{equation} \label{eq:CondCMP}
H(M_A(t_1)|M_A(t_2))<H(M_A(t_2)|M_A(t_1))
\end{equation}

On the other hand, we know from information theory \cite{Cover} that conditional entropy cannot be negative and hence we have: 
\begin{equation} \label{eq:condneg}
0\leq H(M_A(t_1)|M_A(t_2)) 
\end{equation}

By comparing Inequalities~\ref{eq:CondCMP} and~\ref{eq:condneg}, we can easily conclude that $0<H(M_A(t_2)|M_A(t_1))$ which, based on Definition~\ref{def:macrand} in Section~\ref{Section:Energy}, means that system $A$ shows a macroscopically non-deterministic (random) behavior as time goes from $t_1$ to $t_2$ and the theorem is proved. $\blacksquare$

\begin{thmClairvoyanceApp} [The property of clairvoyance] \label{thm:clairvoyanceApp}
Suppose that $A$ is a closed physical system which consists of two smaller systems $B$ and $C$. Assuming that the following three conditions hold for any time interval $t_1$ to $t_2$ ($t_2 > t_1$): 

\begin{itemize}
\itemsep0em
\item[] 1) The system should conform to the second law of thermodynamics, hence we have $\bar{S}_A(t_2) \geq \bar{S}_A(t_1)$.
\item[] 2) System B does not lose any information at the macroscopic level (i.e., it possesses reversibility), hence we have $H(M_B(t_1)|M_B(t_2))=0$. 
\item[] 3) System C has no change in its information content, i.e., does not lose or gain any information. Therefore, we have $H(M_C(t_1)|M_C(t_2))=0$ and $H(M_C(t_2)|M_C(t_1))=0$. 
\end{itemize}

\noindent then during the time interval $t_1$ to $t_2$, we have: 

Part I: If system $B$ has macroscopically non-deterministic behavior (i.e., behaves randomly), which means $H(M_B(t_2)|M_B(t_1))\neq 0$, then it must gain information from system $C$, i.e., we have: 
\begin{equation}\label{eq:ConclusionApp} 
I(M_B(t_2);M_C(t_2)) > I(M_B(t_1);M_C(t_1))
\end{equation}

Part II: But if system $B$ behaves completely in a deterministic manner (i.e., does not behave randomly), which means $H(M_B(t_2)|M_B(t_1))=0$, then it gains no information from system $C$, i.e., we have: 
\begin{equation}\label{eq:Conclusion2App} 
I(M_B(t_2);M_C(t_2)) = I(M_B(t_1);M_C(t_1))
\end{equation}
\end{thmClairvoyanceApp}

\noindent \textbf{Proof:} We first consider the case where $H(M_B(t_2)|M_B(t_1))\neq 0$, i.e., when system $B$ behaves randomly. For this case, as conditional entropy cannot be negative \cite{Cover}, we have: 
\begin{equation}\label{eq:MidEq1} 
H(M_B(t_2)|M_B(t_1)) > 0
\end{equation}

By the chain rule for conditional entropy we have $H(X|Y)=H(X,Y)-H(Y)$ \cite{Cover}, therefore Inequality~\ref{eq:MidEq1} can be rewritten as: 
\begin{equation}\label{eq:MidEq2} 
H(M_B(t_2),M_B(t_1))-H(M_B(t_1)) > 0 
\end{equation}

Again, by the use of the chain rule, the foregoing inequality can be written as: 
\begin{equation}\label{eq:MidEq3} 
H(M_B(t_1)|M_B(t_2))+H(M_B(t_2))-H(M_B(t_1)) > 0
\end{equation}

\noindent but we know from hypothesis~2 that $H(M_B(t_1)|M_B(t_2))=0$, therefore Inequality~\ref{eq:MidEq3} can be simplified to: 
\begin{equation}\label{eq:MidEq4} 
H(M_B(t_2))-H(M_B(t_1)) > 0
\end{equation}

This inequality will be used shortly in the proof, but now we turn to hypothesis~1, which states that the second law of thermodynamics is inviolable. We know from Theorem~\ref{thm:micmac} in Section~\ref{Section:Information} that: 
\begin{equation}\label{eq:MidEq5} 
H(M_A(t_2))+\frac{1}{k_B \ln{2}}\bar{S}_A(t_2)=H(M_A(t_1))+\frac{1}{k_B \ln{2}}\bar{S}_A(t_1)
\end{equation}

From this equation and from the second law, i.e., $\bar{S}_A(t_2) \geq \bar{S}_A(t_1)$, we conclude that: 
\begin{equation}\label{eq:MidEq6} 
H(M_A(t_2)) - H(M_A(t_1)) \leq 0
\end{equation}

From Inequalities~\ref{eq:MidEq4} and~\ref{eq:MidEq6}, we can easily conclude that: 
\begin{equation}\label{eq:MidEq7} 
H(M_B(t_2))-H(M_B(t_1)) > H(M_A(t_2)) - H(M_A(t_1))
\end{equation}

As system $A$ consists of two parts $B$ and $C$, we have $H(M_A(t))=H(M_B(t),M_C(t))$. Therefore, we can rewrite Inequality~\ref{eq:MidEq7} as: 
\begin{equation}\label{eq:MidEq8} 
H(M_B(t_2))-H(M_B(t_2),M_C(t_2)) > H(M_B(t_1))-H(M_B(t_1),M_C(t_1))
\end{equation}

For system $C$, using the chain rule \cite{Cover}, we can write: 
\begin{equation}\label{eq:MidEq9} 
H(M_C(t_1),M_C(t_2))=H(M_C(t_1))+H(M_C(t_2)|M_C(t_1))=H(M_C(t_2))+ H(M_C(t_1)|M_C(t_2))
\end{equation}
However, we know from hypothesis~3 that $H(M_C(t_2)|M_C(t_1))=0$ and $H(M_C(t_1)|M_C(t_2))=0$. Therefore, we conclude from Eq.~\ref{eq:MidEq9} that: 
\begin{equation}\label{eq:MidEq10} 
H(M_C(t_1))=H(M_C(t_2))
\end{equation}

Now, by adding the sides of Eq.~\ref{eq:MidEq10} to the sides of Inequality~\ref{eq:MidEq8} we obtain: 
\begin{equation}\label{eq:MidEq11} 
H(M_B(t_2))+H(M_C(t_2))-H(M_B(t_2),M_C(t_2)) > H(M_B(t_1))+H(M_C(t_1))-H(M_B(t_1),M_C(t_1))
\end{equation}

We know from information theory that $I(X;Y)=H(X)+H(Y)-H(X,Y)$ \cite{Cover}, therefore Inequality~\ref{eq:MidEq11} can be written as: 
\begin{equation}\label{eq:FinalInEq1} 
I(M_B(t_2);M_C(t_2)) > I(M_B(t_1);M_C(t_1))
\end{equation}

\noindent and Part I of the theorem is proved. To prove Part II, first note that, as we have already proved, since $H(M_C(t_2)|M_C(t_1))=0$ and $H(M_C(t_1)|M_C(t_2))=0$ (hypothesis 3), we have $H(M_C(t_1))=H(M_C(t_2))$ (Eq.~\ref{eq:MidEq10}). But we also have $H(M_B(t_2)|M_B(t_1))=0$ (the assumption of Part II) and $H(M_B(t_1)|M_B(t_2))=0$ (hypothesis~2), therefore we must similarly have:
\begin{equation}\label{eq:MidEq12} 
H(M_B(t_1))=H(M_B(t_2))
\end{equation}

Also, based on Lemma~3, we can conclude from $H(M_C(t_2)|M_C(t_1))=0$ (hypothesis~3) and $H(M_B(t_2)|M_B(t_1))=0$ (the assumption of Part II) that: 
\begin{equation}\label{eq:MidEq13} 
H(M_B(t_1), M_C(t_1), M_B(t_2), M_C(t_2))= H(M_B(t_1), M_C(t_1))
\end{equation}

Again, based on Lemma~3, we can conclude from $H(M_C(t_1)|M_C(t_2))=0$ (hypothesis~3) and $H(M_B(t_1)|M_B(t_2))=0$ (hypothesis~2) that: 
\begin{equation}\label{eq:MidEq14} 
H(M_B(t_1), M_C(t_1), M_B(t_2), M_C(t_2))= H(M_B(t_2), M_C(t_2))
\end{equation}
By comparing Eqs.~\ref{eq:MidEq13} and~\ref{eq:MidEq14}, we can conclude that:
\begin{equation}\label{eq:MidEq15} 
H(M_B(t_1), M_C(t_1))= H(M_B(t_2), M_C(t_2))
\end{equation}

Now, from Eqs.~\ref{eq:MidEq10}, \ref{eq:MidEq12}, and~\ref{eq:MidEq15}, we can easily conclude that:
\begin{equation}\label{eq:MidEq16} 
H(M_B(t_1))+H(M_C(t_1))-H(M_B(t_1), M_C(t_1))=H(M_B(t_2))+H(M_C(t_2))-H(M_B(t_2), M_C(t_2))
\end{equation}

We know from information theory that $I(X;Y)=H(X)+H(Y)-H(X,Y)$ \cite{Cover}, therefore Eq.~\ref{eq:MidEq16} can be written as: 
\begin{equation}\label{ eq:FinalInEq2} 
I(M_B(t_1); M_C(t_1))= I(M_B(t_2); M_C(t_2))
\end{equation}

\noindent and the theorem is proved. $\blacksquare$

\begin{thmindiscernibilityApp} [The property of indiscernibility] \label{thm:IndiscernibilityApp}
 
Suppose that A and B are two separate physical systems, i.e., neither of them is a part or component of the other one, then we have: 
\begin{equation}\label{eq: IndiscernibilityApp} 
H(\mathfrak{M}_A(t)|\mathfrak{M}_B(t))= H(\mathfrak{M}_A(t)| M_B(t))
\end{equation}

\end{thmindiscernibilityApp}

\noindent \textbf{Proof:} The theorem is proved as a conclusion from the definitions of perfect microstate and perfect macrostate (Definitions~\ref{def:micro} and~\ref{def:macro} in Section~\ref{Section:Information}). Suppose that system $A$ has some information from system $B$. Based on Definition~\ref{def:micro}, this means that the microstate $\mathfrak{M}_A(t)$ bears some information which is also borne by the microstate  $\mathfrak{M}_B(t)$. Then, $H(\mathfrak{M}_A(t)|\mathfrak{M}_B(t))$ is the information that system $A$ still bears after excluding that part of the information which is also borne by system $B$ (See textbooks on information theory, e.g., \cite{Cover}). 
On the other hand, all the information that system $A$ has from system $B$ must be part of system $B$\textquoteright s perceptible information, otherwise it could not have been perceived by system $A$. This means that if, instead of excluding all the information of system $B$ from the information of system $A$, we only exclude the perceptible information of system $B$ from the information of system $A$, we must have the same result. Therefore, noting that the perfect macrostate of a system represents all the perceptible information of the system (Definition~\ref{def:macro}), we have $H(\mathfrak{M}_A(t)| \mathfrak{M}_B(t))= H(\mathfrak{M}_A(t)|M_B(t))$ and the theorem is proved. $\blacksquare$

\begin{thmMutualApp} [Mutual information between two physical systems] \label{thm:MutualApp}  

Let $A$ and $B$ be two physical systems. Then based on the property of indiscernibility as the only assumption, we have:
\begin{equation}\label{eq:MutualApp} 
I(\mathfrak{M}_A(t);\mathfrak{M}_B(t))=I(\mathfrak{M}_A(t);M _B(t))=I(M_A(t);\mathfrak{M}_B(t))=I(M_A(t);M_B(t))
\end{equation}
\end{thmMutualApp} 
\noindent \textbf{Proof:} We know from information theory \cite{Cover} that we have $I(X,Y)=H(X)-H(X|Y)$ and hence we can write: 
\begin{equation}\label{eq:mutual1}
I(\mathfrak{M}_A(t);\mathfrak{M}_B(t))=H(\mathfrak{M}_A(t))-H(\mathfrak{M}_A(t)|\mathfrak{M}_B(t))
\end{equation}

Based on the property of indiscernibility, we can substitute $H(\mathfrak{M}_A(t)|M_B(t))$ for $H(\mathfrak{M}_A(t)|\mathfrak{M}_B(t))$ and hence Eq.~\ref{eq:mutual1} can be rewritten as: 
\begin{equation}\label{eq:mutual2}
I(\mathfrak{M}_A(t);\mathfrak{M}_B(t))=H(\mathfrak{M}_A(t))-H(\mathfrak{M}_A(t)|M_B(t))
\end{equation}

Based on the equation $I(X,Y)=H(X)-H(X|Y)$ \cite{Cover}, the right hand side of Eq.~\ref{eq:mutual2} can be substituted by $I(\mathfrak{M}_A(t);M_B(t))$, therefore we have: 
\begin{equation}\label{eq:mutualPart1}
I(\mathfrak{M}_A(t);\mathfrak{M}_B(t))=I(\mathfrak{M}_A(t);M_B(t))
\end{equation}

\noindent and one of the equations of the theorem is proved. Due to symmetry, we also have: 
\begin{equation}\label{eq:mutualPart2}
I(\mathfrak{M}_A(t);\mathfrak{M}_B(t))=I(M_A(t);\mathfrak{M}_B(t))
\end{equation}

Therefore, the only remaining part that we need to prove is $I(\mathfrak{M}_A(t);\mathfrak{M}_B(t))=I(M_A(t);M_B(t))$. We prove this by first proving $I(\mathfrak{M}_A(t);\mathfrak{M}_B(t)) \leq I(M_A(t);M_B(t))$ and then proving $I(\mathfrak{M}_A(t);\mathfrak{M}_B(t)) \geq I(M_A(t);M_B(t))$. 

We can use the equation $I(A;B)=H(A)+H(B)-H(A,B)$ \cite{Cover} to expand Eq.~\ref{eq:mutualPart1} as follows:
\begin{equation}\label{eq:Part1Expand}
H(\mathfrak{M}_A(t))+H(\mathfrak{M}_B(t))-H(\mathfrak{M}_A(t),\mathfrak{M}_B(t))=H(\mathfrak{M}_A(t))+H(M_B(t))-H(\mathfrak{M}_A(t),M_B(t))
\end{equation}

Since $M_B(t)$ is a function of $\mathfrak{M}_B(t)$ (Axiom~\ref{ax:mic}) and based on Lemma~\ref{lm:removeREDApp}, we have $H(\mathfrak{M}_B(t))=H(\mathfrak{M}_B(t),M_B(t))$ and $H(\mathfrak{M}_A(t),\mathfrak{M}_B(t))=H(\mathfrak{M}_A(t),\mathfrak{M}_B(t),M_B(t))$. If we substitute these into Eq.~\ref{eq:Part1Expand} and cancel $H(\mathfrak{M}_A(t))$ from both sides of the equation, we obtain:
\begin{equation}\label{eq:Part1Expand2}
H(\mathfrak{M}_B(t),M_B(t))-H(\mathfrak{M}_A(t),\mathfrak{M}_B(t),M_B(t))=H(M_B(t))-H(\mathfrak{M}_A(t),M_B(t))
\end{equation}

Based on the chain rule equation $H(X,Y)=H(X)+H(Y|X)$, Eq.~\ref{eq:Part1Expand2} can be rephrased as:
\begin{equation}\label{eq:Temp1}
H(\mathfrak{M}_B(t)|M_B(t))=H(\mathfrak{M}_B(t)|\mathfrak{M}_A(t),M_B(t))
\end{equation}

In a similar manner and due to symmetry, we can show:
\begin{equation}\label{eq:Temp2}
H(\mathfrak{M}_A(t)|M_A(t))=H(\mathfrak{M}_A(t)|\mathfrak{M}_B(t),M_A(t))
\end{equation}

Now we turn to developing an inequality using these two equations, i.e., Eqs.~\ref{eq:Temp1} and~\ref{eq:Temp2}. We know from Lemma~\ref{lm:MoreCond} that conditioning on more variables leads to more reduction in information, so we can write:
\begin{equation}\label{eq:IneqNoI}
H(\mathfrak{M}_B(t)|M_A(t),M_B(t)) \geq H(\mathfrak{M}_B(t)|\mathfrak{M}_A(t),M_A(t),M_B(t))
\end{equation}

\noindent where the right hand side of the inequality is conditioned on one more variable as compared to the left hand side. Since $M_A(t)$ is a function of $\mathfrak{M}_A(t)$ (Axiom~\ref{ax:mic}), $M_A(t)$ can be removed from the right hand side of Inequality~\ref{eq:IneqNoI}, obtaining:
\begin{equation}\label{eq:IneqNoI2}
H(\mathfrak{M}_B(t)|M_A(t),M_B(t)) \geq H(\mathfrak{M}_B(t)|\mathfrak{M}_A(t),M_B(t))
\end{equation}

Now, we can use Eq.~\ref{eq:Temp1} to substitute for the right hand side of this inequality, and we obtain:
\begin{equation}\label{eq:IneqNoI2_1}
H(\mathfrak{M}_B(t)|M_A(t),M_B(t)) \geq H(\mathfrak{M}_B(t)|M_B(t))
\end{equation}

Based on the chain rule equation $H(X,Y)=H(X)+H(Y|X)$, we can expand the left hand side of Inequality~\ref{eq:IneqNoI2_1}, obtaining: 
\begin{equation}\label{eq:IneqNoI3}
H(\mathfrak{M}_B(t),M_A(t),M_B(t))-H(M_A(t),M_B(t)) \geq H(\mathfrak{M}_B(t)|M_B(t))
\end{equation}

Since $M_B(t)$ is a function of $\mathfrak{M}_B(t)$ (Axiom~\ref{ax:mic}), $M_B(t)$ can be removed from the left hand side of this inequality. Also, we can add $H(\mathfrak{M}_A(t),\mathfrak{M}_B(t))-H(\mathfrak{M}_A(t),\mathfrak{M}_B(t))$ to the left hand side as it is equal to zero, obtaining:
\begin{equation}\label{eq:IneqNoI4}
\begin{split}
[H(\mathfrak{M}_B(t),M_A(t))-H(\mathfrak{M}_A(t),\mathfrak{M}_B(t))]+[H(\mathfrak{M}_A(t),\mathfrak{M}_B(t))-H(M_A(t),M_B(t))] \geq \\ H(\mathfrak{M}_B(t)|M_B(t))
\end{split}
\end{equation}

As $M_A(t)$ is a function of $\mathfrak{M}_A(t)$ (Axiom~\ref{ax:mic}), based on Lemma~\ref{lm:removeREDApp} we have $H(\mathfrak{M}_A(t),\mathfrak{M}_B(t))=H(\mathfrak{M}_A(t),\mathfrak{M}_B(t),M_A(t))$. Therefore, based on the chain rule equation $H(A,B)=H(A)+H(B|A)$, we have:
\begin{equation}\label{eq:TEMP3}
[H(\mathfrak{M}_B(t),M_A(t))-H(\mathfrak{M}_A(t),\mathfrak{M}_B(t))]=-H(\mathfrak{M}_A(t)|\mathfrak{M}_B(t),M_A(t))
\end{equation}

Using Eq.~\ref{eq:Temp2}, the foregoing equation can be written as:
\begin{equation}\label{eq:TEMP4}
[H(\mathfrak{M}_B(t),M_A(t))-H(\mathfrak{M}_A(t),\mathfrak{M}_B(t))]=-H(\mathfrak{M}_A(t)|M_A(t))
\end{equation}

Now, using Eq.~\ref{eq:TEMP4}, Inequality~\ref{eq:IneqNoI4} can be simplified, obtaining: 
\begin{equation}\label{eq:IneqNoI5}
H(\mathfrak{M}_A(t),\mathfrak{M}_B(t))-H(M_A(t),M_B(t)) \geq H(\mathfrak{M}_B(t)|M_B(t))+H(\mathfrak{M}_A(t)|M_A(t))
\end{equation}

Using the chain rule equation $H(A,B)=H(A|B)+H(B)$, this inequality can be rephrased as:
\begin{equation}\label{eq:IneqNoI6}
\begin{split}
H(\mathfrak{M}_A(t),\mathfrak{M}_B(t))-H(M_A(t),M_B(t)) \geq \\ H(\mathfrak{M}_B(t),M_B(t))-H(M_B(t))+H(\mathfrak{M}_A(t),M_A(t))-H(M_A(t))
\end{split}
\end{equation}

Since $M_A(t)$ is a function of $\mathfrak{M}_A(t)$ (Axiom~\ref{ax:mic}), $M_A(t)$ can be removed from $H(\mathfrak{M}_A(t),M_A(t))$ (Lemma~\ref{lm:removeREDApp}). The same can be done for $M_B(t)$ and $H(\mathfrak{M}_B(t),M_B(t))$. Therefore, Inequality~\ref{eq:IneqNoI6} can be written as:
\begin{equation}\label{eq:IneqNoI7}
H(M_A(t))+H(M_B(t))-H(M_A(t),M_B(t)) \geq H(\mathfrak{M}_A(t))+H(\mathfrak{M}_B(t))-H(\mathfrak{M}_A(t),\mathfrak{M}_B(t))
\end{equation}

On the other hand, as we have $I(A;B)=H(A)+H(B)-H(A,B)$ \cite{Cover}, the foregoing inequality can be written as: 
\begin{equation}\label{eq:IneqNoI8}
I(M_A(t);M_B(t)) \geq I(\mathfrak{M}_A(t);\mathfrak{M}_B(t))
\end{equation}

To prove that we also have $I(M_A(t);M_B(t)) \leq I(\mathfrak{M}_A(t);\mathfrak{M}_B(t))$, we begin with the following equation, which is valid because of Lemma~\ref{lm:removeREDApp}, and because $M_A(t)$ and $M_B(t)$ are functions of $\mathfrak{M}_A(t)$ and $\mathfrak{M}_B(t)$ respectively (Axiom~\ref{ax:mic}). 
\begin{equation}\label{eq:MUTMUT} 
I(\mathfrak{M}_A(t); \mathfrak{M}_B(t))=I(\mathfrak{M}_A(t),M_A(t); \mathfrak{M}_B(t),M_B(t))
\end{equation}

On the other hand, for any three variables $X$, $Y$, and $Z$, the following inequality holds:
\begin{equation}\label{eq:TAUTIneq} 
I(X,Y;Z) \geq I(X;Z)
\end{equation}

This is due to a rule in information theory, called the chain rule for mutual information, which is
\cite{Cover}:
\begin{equation}\label{eq:MUTChainRule} 
I(X,Y;Z)=I(X;Z)+I(Y;Z|X)
\end{equation}

Conditional mutual information cannot be negative \cite{Cover}, which means that the term $I(Y;Z|X)$ in Eq.~\ref{eq:MUTChainRule} cannot be negative and hence Eq.~\ref{eq:MUTChainRule} implies Inequality~\ref{eq:TAUTIneq}. Based on Inequality~\ref{eq:TAUTIneq}, we have:
\begin{equation}\label{eq:TwoIneq} 
I(\mathfrak{M}_A(t),M_A(t); \mathfrak{M}_B(t),M_B(t)) \geq 
I(M_A(t); \mathfrak{M}_B(t),M_B(t)) \geq
I(M_A(t); M_B(t))
\end{equation}

But we know from Eq.~\ref{eq:MUTMUT} that $I(\mathfrak{M}_A(t); \mathfrak{M}_B(t))=I(\mathfrak{M}_A(t),M_A(t); \mathfrak{M}_B(t),M_B(t))$, therefore we conclude from Inequality~\ref{eq:TwoIneq} that:
\begin{equation}\label{eq:IneqFinal2} 
I(\mathfrak{M}_A(t); \mathfrak{M}_B(t)) \geq I(M_A(t); M_B(t))
\end{equation}

From Inequalities~\ref{eq:IneqNoI8} and~\ref{eq:IneqFinal2} we conclude:
\begin{equation}\label{eq:EqFinal} 
I(\mathfrak{M}_A(t); \mathfrak{M}_B(t)) = I(M_A(t); M_B(t))
\end{equation}

We have proved all the four equations of the theorem and hence the theorem is proved. $\blacksquare$
\begin{thmAdditivityApp} [The property of additivity] \label{thm:additivityApp}
Let $A$ and $B$ be two physical systems. Then by only assuming the property of indiscernibility and without any other assumption, we have: 
\begin{equation}\label{eq:additivityApp} 
H(\mathfrak{M} _A(t), \mathfrak{M} _B(t)|M_A(t), M_B(t))= H(\mathfrak{M} _A(t)|M_A(t))+ H(\mathfrak{M} _B(t)|M_B(t))
\end{equation}
\end{thmAdditivityApp}

\noindent \textbf{Proof:} We have proved in Theorem~\ref{thm:MutualApp} that by assuming the property of indiscernibility, we have:
\begin{equation}\label{eq:MutualMain} 
I(\mathfrak{M}_A(t);\mathfrak{M}_B(t))=I(M_A(t);M_B(t))
\end{equation}

Now we prove that Eq.~\ref{eq:MutualMain} is indeed equivalent to the property of additivity, i.e., Eq.~\ref{eq:additivityApp}, thereby proving the theorem. To do this, we use the equation $I(X,Y)=H(X)+H(Y)-H(X,Y)$ \cite{Cover} from information theory to expand Eq.~\ref{eq:MutualMain} as follows:
\begin{equation}\label{eq:MutualMainExpand} 
H(\mathfrak{M}_A(t))+H(\mathfrak{M}_B(t))-H(\mathfrak{M}_A(t),\mathfrak{M}_B(t))= H(M_A(t))+H(M_B(t))-H(M_A(t), M_B(t))
\end{equation}

Using elementary algebra, this equation can be rewritten as:
\begin{equation}\label{eq:ExpandReWr} 
H(\mathfrak{M}_A(t),\mathfrak{M}_B(t))-H(M_A(t),M_B(t)) = H(\mathfrak{M}_A(t))-H(M_A(t))+H(\mathfrak{M}_B(t))-H(M_B(t))
\end{equation}

On the other hand, since $M_A(t)$ and $M_B(t)$ are functions of $\mathfrak{M}_A(t)$ and $\mathfrak{M}_B(t)$ respectively (Axiom~\ref{ax:mic}), and due to Lemma~\ref{lm:removeREDApp}, Eq.~\ref{eq:ExpandReWr} can be rewritten as:
\begin{equation} \label{eq:WtRD} 
\begin{split}
H(\mathfrak{M}_A(t),\mathfrak{M}_B(t),M_A(t),M_B(t))-H(M_A(t),M_B(t)) = \\ H(\mathfrak{M}_A(t), M_A(t))-H(M_A(t))+H(\mathfrak{M}_B(t), M_B(t))-H(M_B(t))
\end{split}
\end{equation}

Now using the chain rule equation $H(X,Y)=H(X)+H(Y|X)$ \cite{Cover}, Eq.~\ref{eq:WtRD} can be rephrased as:
\begin{equation}\label{eq:ExpandRph} 
H(\mathfrak{M}_A(t),\mathfrak{M}_B(t)|M_A(t),M_B(t)) = H(\mathfrak{M}_A(t)| M_A(t))+H(\mathfrak{M}_B(t)|M_B(t))
\end{equation}

\noindent and the theorem is proved. $\blacksquare$
\\~\\~\\

\noindent {\Large {Appendix 3: Change in thermodynamic entropy due to solving the satisfiability problem}}

In this appendix, we consider the reversible computer of Section~\ref{Section:Daemon}, which is used for solving the satisfiability problem with the assumptions given in that section. As mentioned in Section~\ref{Section:Daemon}, for this computer, the thermodynamic entropy change is given by $\Delta \bar{S}=-k_B \ln{2} \cdot H(O|I)$. Hence, here we turn to calculating $H(O|I)$. From information theory \cite{Cover}, we know that:
\begin{equation}\label{eq:condentApp3} 
H(O|I)=\underset{i}{\mathbb{E}}[H(O|I=i)]=\sum_{i\in D_I} [P(I=i)H(O|I=i)]
\end{equation}

\noindent where $D_I$ is the universe of $I$, i.e., the set of all possible inputs to the computer, and $H(O|I=i)$ is:
\begin{equation}\label{eq:condent2App3} 
H(O|I=i)=-\sum_{j\in D_O} [P(O=j|I=i)\log_2 P(O=j|I=i)]
\end{equation}

We know that $i$ in Eqs.~\ref{eq:condentApp3} and~\ref{eq:condent2App3} is a Boolean expression as it is the input to a computer that solves the satisfiability problem. Suppose that $i$ is an expression with $k$ different possible satisfying variable assignments. Then, based on the uniform distribution assumption in Section~\ref{Section:Daemon}, any of these satisfying assignments might appear at the output with the probability $\frac{1}{k}$ and other (unsatisfying) assignments cannot appear at the output, i.e., their probability is $0$. Therefore, in this case, Eq.~\ref{eq:condent2App3} can be rewritten as:
\begin{equation}\label{eq:condent3App3} 
H(O|I=i)=-\sum_{j=1}^{k} \frac{1}{k} \log_2 \frac{1}{k}=\log_2 k \quad \textrm{when the expression } i \textrm{ has } k \textrm{ satisfying assignments.} 
\end{equation}

Now to obtain $H(O|I)$ we need to substitute for $H(O|I=i)$ from Eq.~\ref{eq:condent3App3} into Eq.~\ref{eq:condentApp3}. To do this, we first note that the number of $n$-variable Boolean functions is $2^{2^{n}}$, and also the number of functions with $k$ satisfying assignments is ${2^{n}\choose k}$\footnote{ These can be easily understood by noting that any Boolean function can be defined by a truth table. An $n$-variable function has a truth table with $2^{n}$ rows and for each row the function can have one of the two possible values True or False. Therefore, there are $2^{2^{n}}$ different $n$-variable Boolean functions. Also, a Boolean function that has $k$ satisfying assignments is True for exactly $k$ rows of its truth table and is False for all the other rows, hence there are ${2^{n}\choose k}$ functions that have $k$ satisfying assignments.}. Therefore, based on the uniform distribution assumption in Section~\ref{Section:Daemon}, any Boolean function might be given as an input expression with the probability $\frac{1}{2^{2^{n}}}$  and also there are ${2^{n}\choose k}$  functions that have the same value for $H(O|I=i)$, which is $\log_2 k$. Therefore, Eq.~\ref{eq:condentApp3} can be rewritten as:
\begin{equation}\label{eq:HOI_NPApp} 
H(O|I)= \frac{1}{2^{2^{n}}} \sum_{k=2}^{2^{n}} {2^{n}\choose k}\cdot \log_2 k
\end{equation}
\\~\\~\\



\end{document}